\documentclass[11pt]{article}
 
\usepackage{graphicx}
\usepackage{subcaption}
\usepackage{float}
\usepackage[dvipsnames]{xcolor}
\usepackage{multicol} 

\usepackage{color}
\usepackage{latexsym}

\definecolor{rosso}{cmyk}{0,1,1,0.4}
\definecolor{rossos}{cmyk}{0,1,1,0.55}
\definecolor{rossoc}{cmyk}{0,0.5,1,0.2}
\definecolor{blu}{cmyk}{1,1,0,0.3}
\definecolor{blus}{cmyk}{1,1,0,0.6}
\definecolor{blucc}{cmyk}{1,0.4,0.2,0}
\definecolor{viola}{cmyk}{0,1,0,0.6}
\definecolor{viola2}{cmyk}{0,1,0.2,0.6}
\definecolor{verde}{cmyk}{0.92,0,0.59,0.25}
\definecolor{verdec}{cmyk}{0.92,0,0.59,0.15}
\definecolor{verdes}{cmyk}{0.92,0,0.59,0.4}
\font\tenrsfs=rsfs10 at 12pt
\font\sevenrsfs=rsfs7
\font\fiversfs=rsfs5
\newfam\rsfsfam 
\textfont\rsfsfam=\tenrsfs
\scriptfont\rsfsfam=\sevenrsfs
\scriptscriptfont\rsfsfam=\fiversfs
\def\mathscr#1{{\fam\rsfsfam\relax#1}}

\def\circa#1{\,\raise.3ex\hbox{$#1$\kern-.75em\lower1ex\hbox{$\sim$}}\,}

\usepackage{amsmath,latexsym,amssymb,color,hyperref,graphicx}

\newcommand{\be}{\begin{equation}}
\newcommand{\ee}{\end{equation}}
\newcommand{\bea}{\begin{eqnarray}}
\newcommand{\ena}{\end{eqnarray}}
\newcommand{\no}{\noindent}
\newcommand{\nb}{\nonumber}

\newcommand\m{\ensuremath{\mu}}

\newcommand{\de}{\partial}
\newcommand{\ha}{\frac{1}{2}}

\newcommand{\ea}{\end{eqnarray}}
\newcommand{\plm}{M_{\text{Pl}}} 
 
\makeatletter
\def\ps@mine{%
    \def\@oddfoot{\hfil\thepage\hfil}\let\@evenfoot\@oddfoot
    \let\@oddhead\@evenhead%
    \let\@mkboth\@gobbletwo
    \let\sectionmark\@gobble
    \let\subsectionmark\@gobble
    }
\renewcommand\section{\@startsection {section}{1}{\z@}%
                                   {-3.5ex \@plus -1ex \@minus -.2ex}%
                                   {2ex \@plus.2ex}%
                                   {\normalfont\large\sffamily\bfseries}}
\renewcommand\subsection{\@startsection {subsection}{1}{\z@}%
                                   {-3.5ex \@plus -1ex \@minus -.2ex}%
                                   {2ex \@plus.2ex}%
                                   {\normalfont\sffamily\bfseries}}
\makeatother

\numberwithin{equation}{section}

\begin{document}

\thispagestyle{empty}
\vspace*{-2.5cm}
\begin{minipage}{.45\linewidth}

\end{minipage}
\vspace{2.5cm}

\begin{center}
  {\huge\sffamily\bfseries Boosting GWs in Supersolid Inflation
  }

  \end{center}
 
 \vspace{0.5cm}
 
 \begin{center} 
 {\sffamily\bfseries \large  Marco Celoria}$^{b}$,
 {\sffamily\bfseries \large  Denis Comelli}$^{c}$,
 {\sffamily\bfseries \large  Luigi Pilo}$^{d,e}$,
  {\sffamily\bfseries \large Rocco Rollo}$^{a,d}$\\[2ex]
$^a$ Gran Sasso Science Institute (GSSI)\\Viale Francesco Crispi 7, I-67100 L'Aquila, Italy\\\vspace{0.3cm}
$^b$ ICTP, International Centre for Theoretical Physics Strada Costiera 11, 34151, Trieste, Italy \\\vspace{0.3cm}
$^c$ INFN, Sezione di Ferrara, I-44122 Ferrara, Italy \\\vspace{0.3cm}
$^d$INFN, Laboratori Nazionali del Gran Sasso, I-67010 Assergi, Italy\\\vspace{0.3cm}
$^e$Dipartimento di Ingegneria e Scienze dell'Informazione e Matematica, Universit\`a degli Studi dell'Aquila,  I-67010 L'Aquila, Italy\\\vspace{0.3cm}
 {\tt mceloria@ictp.it}, {\tt comelli@fe.infn.it}, {\tt luigi.pilo@aquila.infn.it}, {\tt rocco.rollo@gssi.it},
\end{center}

\vspace{0.7cm}

\begin{center}
{\small \today}
\end{center}

\vspace{0.7cm}

\abstract{
 Inflation driven by a generic self-gravitating medium is an
interesting alternative to study the impact of spontaneous spacetime
symmetry breaking during a quasi de-Sitter phase, in particular 
the 4-dimensional diffeomorphism invariance of
GR is spontaneously broken down to $ISO(3)$.
The effective description is based on  four scalar fields
  that  describe the excitations of a supersolid. 
There are two phonon-like propagating scalar degrees of freedom
that mix non-trivially both at early and late times and, after exiting
  the horizon, give rise to non-trivial correlations among the different
  scalar power spectra.
  The non-linear structure of the theory
  allows  a secondary gravitational waves production during inflation, efficient enough  to saturate the
present experimental bound and with a blue-tilted spectral index.} 
\clearpage

\section{Introduction}

Inflation is the most compelling way to solve the drawbacks of the 
hot big bang  model and simultaneously generate the seed of the primordial
perturbations to be used as initial conditions for the
latter stages of Universe's evolution. The simplest class of models is single
clock inflation, where time diffeomorphisms are non-linearly realized,   
whose predictions are largely independent on how the
Universe is reheated. Indeed,  according  to the Weinberg
theorem on cosmological perturbations~\cite{Weinberg:2003sw,Weinberg:2008zzc}, 
at large scales and under mild assumptions, the curvature
perturbations of the constant density hypersurfaces $\zeta$,   
or equivalently the comoving curvature ${\cal R}$, are conserved and
can be used to set the primordial initial conditions for the scalar
sector at the beginning of radiation domination. 
The situation is different for models characterized by different symmetry breaking patterns, 
featuring  more degrees of freedom for which the
Weinberg theorem does not hold.  In this case,
${\cal R}$ and $\zeta$ are not conserved
  and  ${\cal R} \neq \zeta$  at superhorizon scales; 
  thus, the details of reheating have to be taken into account
\cite{Kinney:2005vj,Namjoo:2012aa,Motohashi:2014ppa,Akhshik:2015nfa,Celoria:2017xos}. 
That is exactly what happens when a fluid~\cite{Chen:2013kta} or solid~\cite{Endlich:2012pz}
drives inflation. 
In this work, we present an effective field theory (EFT)
description suitable to describe the complete breaking of spacetime
diffeomorphisms during inflation by using the minimal set of four scalar
fields  $\{\varphi^A \, , \;  A=0,1,2,3 \}$ sporting a suitable set of internal symmetries. 
As a matter of fact, $\varphi^A$ can also be interpreted as the coordinates of a self-gravitating  non-dissipative medium~\cite{Matarrese:1984zw,Dubovsky:2005xd,Dubovsky:2011sj,Ballesteros:2012kv,Ballesteros:2016kdx,Celoria:2017bbh} that in our case
is a supersolid. A complete analysis of the linear evolution of scalar
and tensor modes,  together  with the computation of
the corresponding power spectra is given. In addition,
we consider the secondary production of gravitational waves (GWs)
during inflation, exploiting the cubic vertex of the theory involving
a tensor and two scalars, saturating the experimental bound set by Planck without upsetting the scalar 3-point function. 
The secondary production can give a blue tilt to the spectral index, an important feature for the direct detection of the stochastic GWs background produced during inflation. 
A detailed analysis of non-Gaussianity can be found in a companion paper~\cite{NGSS}.  

The outline of the paper is the following.
In section \ref{EFT} we briefly review the effective field theory
description of a supersolid at the leading order in derivates.
In section \ref{cosmology}, the dynamics of the two independent  scalar
perturbations are careful analyzed both at classical and quantum
level, computing the relevant scalar power spectra and  constraining
the parameter region by using  Plank data.    
Section \ref{reheating} is devoted to study, in the instantaneous
reheating approximation, how the seed of primordial perturbations are
transmitted to the radiation dominated phase in a $\Lambda$CDM universe. 
In section \ref{G_W} primary and secondary gravitational waves
production during inflation is considered.  Our conclusions are drawn in section \ref{con}.

\section{Supersolids and Inflation}
\label{EFT}

Several features of inflationary models can be traced back to the
spontaneous symmetry breaking pattern: in single field inflation, the non-trivial time-dependent configuration of the inflaton breaks time reparametrization leaving unbroken the space diffeomorphisms of the $t=$const. hypersurface. 
However, there are other possibilities. For instance, an inflationary model 
where spatial diffeomorphisms are non-linearly realized 
was studied in~\cite{Endlich:2012pz} by working with three scalar fields. 
In a similar fashion, one can consider a more general case in which all
diffeomorphisms are broken by the background configuration of four scalar fields
\be
\varphi^0 = \bar \varphi(t) \, , \qquad \qquad \varphi^i = x^i \,
\label{vac}
\ee
which will be the background configuration for the inflationary phase.
The existence of a spatially homogeneous background is allowed by the
presence of global symmetries of the scalar field action. 
Consider a special multi-field  model of inflation based on four scalar fields $\varphi^A$, 
$A=0\,,1\,,\,2\,,\,3$  with shift symmetry~\cite{Dubovsky:2005xd,Celoria:2017bbh} 
\be 
\varphi^A \rightarrow \varphi^A +
c^A\, ,
\label{shift}
\ee
and  $SO(3)$ internal symmetry
\be
\varphi^a \to {\varphi'}^a= {\cal
  R}^a_b \;\varphi^a \,,  \qquad a= 1,2,3 \qquad \qquad {\cal R}^t \;{\cal R} =1
\, .
\label{inrot}
\ee
The ``vacuum'' configuration (\ref{vac}) has  a residual  global ``diagonal'' $ISO(3)$ 
symmetry. Indeed, a global spatial rotation ${\cal R}^i_j x^j$ can be
absorbed by a corresponding inverse internal transformation of
$\varphi^a$ and the same is true for a global translation $x^i \to x^i +c^i$ thanks to
the shift symmetry  (\ref{shift}).
Among the spacetime scalars shift symmetric operators with a single derivative of $\varphi^A$
\be
 C^{AB} = g^{\mu \nu} \de_\mu \varphi^A \, \de_\nu \varphi^B \;\;
 \qquad A,B=0,1,2,3  \, , 
\ee
one can extract 10  operators invariant under internal $SO(3)$
rotations (\ref{inrot}) 
\be
\begin{split}
& b=\sqrt{\text{Det}\left[\textbf{B}\right]}\, ,\qquad
\varsigma=\sqrt{\text{Det}\left[\textbf{W}\right]}\, ,
\qquad y=u^\mu \, \partial_\mu \varphi^0\,,\qquad
\chi=\sqrt{-C^{00}}\, , \\
&\tau_X= \text{Tr}\left[\textbf{B}\right]\,,\qquad \tau_Y=
\frac{\text{Tr} \left[\textbf{B}^2\right]}{\tau_X{}^2}\,,\qquad
\tau_Z= \frac{\text{Tr} \left[\textbf{B}^3\right]}{\tau_X{}^3}\, ,  \\
&w_X= \text{Tr}\left[\textbf{W}\right]\,,\qquad w_Y= \frac{\text{Tr}
  \left[\textbf{W}^2\right]}{w_X{}^2}\,,\qquad w_Z= \frac{\text{Tr}
  \left[\textbf{W}^3\right]}{w_X{}^3}\, ;
\end{split} 
\ee
where
\be
u^\mu=
-\frac{\epsilon^{\mu\nu\alpha\beta}}{6\;b\;\sqrt{-g}}\,\epsilon_{abc}\;\partial_\mu\varphi^a\,\partial_\nu\varphi^b\,\partial_\beta\varphi^c
\, , \qquad u^2=-1 \, , 
\ee
plays the role of  the medium four-velocity such that $u^\mu\partial_\mu\varphi^a=0$ 
and
\be
B^{ab}=C^{ab}\,, \qquad \qquad W^{ab} = B^{ab} - \frac{C^{0a} \,
  C^{0b}}{C^{00}} \, , \qquad  a=1,\,2,\, 3\, .
\ee
By using the relation $\varsigma =\chi^{-1}  \, b \, y$ and the
Cayley-Hamilton theorem, only $7$ among those operators
are independent. Thus, we arrive at the action
\be 
S=\frac{\plm^2}{2}\, \int dx^4\, \sqrt{-g}\, R+\plm^2 \, \int dx^4 \,
\sqrt{-g}\, U(b,\,y,\,\chi,\,\tau_Y,\,\tau_Z,\, w_Y,\, w_Z)\,,
\label{gact}
\ee
that can be interpreted as the relativistic generalization of the  low-energy effective Lagrangian describing 
homogeneous and isotropic  supersolids at zero-temperature \cite{Son:2005ak, Landry:2019iel}.
Such an action is the most general at leading order in a derivative
expansion compatible with (\ref{shift}) and (\ref{inrot}) and it is rather useful to study systematically the
symmetry breaking pattern of spacetime symmetry during inflation. 

By inspection of the energy momentum tensor (EMT),
the energy density and the  pressure are given by
\bea\label{ed}
&& \rho=\plm^2\, \left( -U+\chi \, U_\chi+y\, U_y \right)  \,,\\
&& p= \plm^2\, \left(U-b\, U_b \right) \, .
\label{pre}
\ea
According to the Noether theorem, there are four conserved currents:
\be
J^\mu_A =2 \, \plm^2\, \frac{\de U}{\de C^{AB}} \nabla^\mu \varphi^B \, ;
\ee
three related to solid configurations that spontaneously break translation invariance and one associated with the superfluid frictionless flow.  
In particular, the  particle number density $n_{sf}$ of the superfluid component can be expressed in terms of the
Noether current $J^0_\mu$ as 
\be 
n_{sf}=-u^\alpha \, J_{\alpha}^0\, ;
\ee
while the density of lattice sites $n_\ell$ is identified as the projection  of the off
shell  conserved current~\footnote{The conservation
  of $J^\mu$, $\nabla^\alpha \, J_\alpha=0$, holds without the use of
  the equations of motion for $\varphi^a$.} $J_\alpha= b \,u_\alpha$ along the
four-velocity $u^\mu$, namely \cite{Son:2005ak}
\be\label{defn}
 n_\ell=-u_\alpha \, J^\alpha = b \, . 
\ee 
This allows us to define the superfluid density   per lattice site $\sigma$ as
\be 
\sigma= \frac{n_{sf}}{n_\ell}= \frac{\plm^2}{b} \left(U_y+\frac{y}{\chi}\,
  U_{\chi}\right)\,.
\label{sigmadef}
\ee
As we will see, at  cosmological level, the perturbations
$\delta\sigma$ generate non-adiabatic contributions (for this reason,
we will regard $\delta\sigma$ in the following as an isocurvature or entropic perturbation).
Similarly, for the rest of the paper we identify $n_\ell$ (\ref{defn}) with the {\it usual} particle density $n$.
\\
While $u^\mu$ represents the 4-velocity of the normal component of the supersolid,
\be
{\cal V}_\mu = \frac{\de_\nu \varphi^0}{\chi}
\label{svel}
\ee
is the  4-velocity (irrotational) of the superfluid component.

One of the key features is that two independent phonon-like excitations
are present. In general, the supersolid perturbation can be written
around a flat space-time as
\be
g_{\mu \nu} =\eta_{\mu \nu} , \qquad \varphi^0 = t + \pi_0 \, , \qquad
\varphi^i=x^i+\partial^i \pi_L+\pi_T^i \,, \qquad \de_i  \pi_T^i=0 \,  .
\ee
At the quadratic level, we have \cite{Celoria:2017hfd}
\begin{equation}
\begin{aligned}
{\cal S}^{(2)} &= \int d^4 x \Big [ \frac{( \hat{M_1} + \bar{\rho}+\bar{p})}{2} \,\de_t\pi_L\, \Delta\,
  \de_t \pi_L +  \left(\hat{M}_3- \hat{M}_2 \right) \pi_L \, \Delta^2
  \, \pi_L+
   \hat{M}_0 \,
  {\pi_0'}^2+ \left(2\,  \hat{M}_4 - \hat{M}_1 \right) \pi_0' \Delta
  \pi_l \nb\\
& - \frac{\hat{M}_1}{2} \pi_0 \, \Delta \, \pi_0+\frac{(\hat{M}_1 + \bar{\rho} +\bar{ p})}{2}\,
  \de_t \pi_T^i \,\de_t \pi_T^i - \frac{\hat{M}_2}{2} \pi_T^i\, \Delta \, \pi_T^i
\Big ] \, ;
\label{quadflat}
\end{aligned}
\end{equation}
where $\bar{\rho}$ and $\bar{p}$ are the background values of the energy
  density and pressure (constant in space and time) while $\Delta =
  \delta_{ij} \de_i \de_j$;  finally, the derivative  of a function
  $f$ with respect to conformal time is denoted by $f'$. The
parameters $\hat{M}_a= \plm^2 \, M_a$ are
proportional to first and second  derivatives of $U$ and are given in appendix
\ref{mass-app}. Notice that the space shift symmetry is crucial to  have
a  homogeneous EMT even if the scalar fields have
non-trivial background values. \\
The properties of the EMT are largely determined by the symmetries of
$U$ as discussed\footnote{In
  ~\cite{Ballesteros:2016kdx,Celoria:2017bbh} the set chosen
  independent operators is different from our choice without changing
  the physics.}  in~\cite{Ballesteros:2016kdx,Celoria:2017bbh}. It is
useful to summarize the main features associated with the presence or
absence of some of the operators (and the related mass
parameters) in the Lagrangian depending on a specific set of internal symmetries.
\begin{itemize}
\item Perfect Fluids: 
\begin{itemize}
\item $U(b):$
only $\{\varphi^a \, , \; \; a=1,2,3 \}$  are present;  the Lagrangian is invariant  under internal volume
preserving diffeomorphisms $V_s\text{Diff}$:
$\varphi^a\to\Psi^a(\varphi^b)$\,,\quad $\det |\partial \Psi^a
/\partial{\varphi^b}|=1$, $a,b=1,2,3$.
\item $U(\chi)$: 
  it is the most general Lagrangian for a perfect irrotational fluid
  with $\varphi^0$ only.
\item
  $U(b,\,y)$: the most general isentropic perfect fluid; the Lagrangian is invariant under $V_s\text{Diff}$ and $\varphi^0 \to \varphi^0 +f(\varphi^a)$. 
\end{itemize}
\item Superfluids 
  $U(b,\,y,\,\chi)$: 
invariant under volume  preserving diffeomorphisms $V_s\text{Diff}$.
\item Solids $U(b,\,\tau_Y,\,\,\tau_Z)$: \footnote{The operator  $\tau_X$ can be written as a function of $(b,\,\tau_Y, \tau_Z)$, so only three operators are independent.} is the
 most general Lagrangian  with only $\{\varphi^a \, , \; \; a=1,2,3 \}$ present.
\item Zero-Temperature Supersolids 
  $U(b, \, y,\,\chi,\,\tau_Y,\,\,\tau_Z, \, w_Y, \, w_Z)$. 
\end{itemize}  
As a remark, in the literature solids are  typically associated with the presence of only three scalar fields $\{\varphi^a \, , a=1,2,3 \}$ and a Lagrangian of the form $U(\tau_X, \tau_Y, \tau_Z)$, see for instance~\cite{Endlich:2012pz}. However, introducing a fourth scalar
  $\varphi^0$ and enforcing the following field dependent shift
  symmetry 
\begin{equation}
\varphi^0 \to \varphi^0+f(\varphi^a)\,,
\end{equation}
the allowed operators are $b$, $\tau_Y$, $\tau_Z$ and $y$ and  the resulting theory $U(b,\,\tau_Y,\,\tau_Z,\,y)$ describes an adiabatic solid in the sense that
the entropic perturbation $\delta \sigma$ is a conserved quantity as  
discussed in~\cite{Celoria:2017bbh,Celoria:2019oiu}.
The term supersolid is reserved to the case where, in
  addition  to the phonons of the solid, also  the entropic
perturbation $\delta \sigma$ propagates. 

A more detailed analysis of thermodynamical properties for general
supersolids is planned for a future work. 
Finally, stability of (\ref{quadflat}) imposes the following
conditions \cite{Celoria:2017hfd}
\be
\hat M_0 >0 \, , \qquad -(\bar{p}+ \bar{\rho} )< \hat{M}_1 < 0 \, , \qquad
\hat M_2 >0\, , \qquad \hat M_2 > \hat M_3 \, .
\label{fstab}
\ee
As we will see, such conditions are necessary for the existence of the Bunch-Davies (BD) vacuum in an inflating phase driven by a supersolid.
During a quasi deSitter period, the most convenient parametrization of the mass term is
 through some $c_i^2$ parameters such that 
\be
\label{par}
M_i\equiv \plm^{-2}\;(1+w) \;\rho\; c_i^2   \qquad i=0,1,2,3,4  \, . 
\ee
(where $w$ is defined in (\ref{eqw})) with the assumption that $c_i^2 $ are slowly varying in time ($ \frac{c_i'}{{\cal H}\, c_i} \ll 1$).

\section{Slow-roll}
\label{cosmology}
Cosmological perturbations in the flat-slice gauge are described by 
\be
\begin{split}
&\varphi^0 =  \bar \varphi(t) + \pi_0 \, , \qquad
\varphi^i=x^i+\partial^i \pi_L+\pi_T^i \,, \qquad \de_i  \pi_T^i=0 \,
, \\
& ds^2=a^2\left[ (2 \, \Psi-1 )\; dt^2 + 2 \, dt \;dx^i \, \de_i F + \delta_{ij} \;dx^i \;dx^j\right] \, .
\end{split}
\label{flatg}
\ee
Perturbations in a generic gauge are discussed in Appendix
\ref{gi-app}.
The background EMT tensor has the form of the one of a
perfect fluid with energy density and pressure given by (\ref{ed}) and
(\ref{pre}) evaluated on FLRW; the conservation of the background EMT is equivalent to~\footnote{In \cite{Bartolo:2015qvr}
  it was set $\bar {\varphi}'=a$  that  leads to a conserved
  background EMT only if $c_b^2=0$. Such a value of $c_b^2$  is rather peculiar, as we will see in what follows. Moreover, the correct implementation of the Stuckelberg trick at the background level requires a
non-trivial background for $\varphi^0$ satisfying (\ref{tbkg}).}
\be
\bar\varphi''={\cal H}\,(1-3\, c_b^2)\;\bar\varphi' \, ,
\label{tbkg}
\ee
where
\be
\label{cb}
c_b^2 =
- \frac{c_4^2}{c_0^2} \, .
\ee
For $c_b^2$ constant in time, we have 
\be
\bar\varphi'=\bar\varphi'_0\;a^{1-3\;c_b^2},\qquad \bar\varphi'_0=const
\ee
Our benchmark values for $c_b^2$ will be $c_b^2=0 $ which gives $
\bar\varphi'=a^{ }\;  \bar\varphi'_0$, and $c_b^2=-1$ leading to $ \bar\varphi'=\bar\varphi'_0\;a^{ -4}$.
Inflation takes place when
\be\label{eqw}
w < - \frac{1}{3} \, , \qquad w = \frac{\bar{p}}{\bar{\rho}} \, .
\ee
We will be mainly  interested in slow-roll (SR) inflation~\footnote{As
  discussed in \cite{Celoria:2019oiu} super SR is also possible; actually when $M_2=0$, as for fluids, this is the only viable regime with small $\epsilon$.} for which the following dynamical parameters are small
\be
\epsilon=1-\frac{{\cal H}'}{{\cal H}^2}= \frac{3}{2}\,(1+w)
\,, \qquad \qquad \eta= \frac{\epsilon'}{\epsilon \, {\cal H}} \ll
1\, .
\label{epsdef}
\ee
Note that in a quasi dS phase, the adiabatic speed of sound is given by
\be
c_s^2 =\frac{p'}{\rho'} = \frac{2}{3} \left(3\,  c_0^2 \,
  c_b^4+c_2^2-3 \, c_3^2\right)=-1+\frac{2}{3}\,
\epsilon-\frac{1}{3}\,\eta \,
{}_{| \text{SR}}\simeq -1.
\label{dsss}
\ee
Both $\Psi$ and $F$ are non-dynamical fields and their linear
equations of motion can be solved in terms of $\pi_0$ and $\pi_L$, at
the leading order in SR and working  in Fourier space, one finds
\be
\begin{split}
  & \Psi= \frac{c_1^2 \,\mathcal{H} \,\epsilon }{\bar{\varphi} '} \pi _0 -\mathcal{H}
  \epsilon \, \left(c_1^2+1\right) \,\pi _L' \, ;\\
  & F={\cal H}\,(2 \, c_0^2\, c_b^2-1) \,\pi_L\,\epsilon-3\, c_1^2 \,k^{-2}\, {\cal H}^2\,\frac{\pi_0}{\bar\varphi'}\,\epsilon+3\, (1+c_1^2)\,{\cal H}^2 \, k^{-2}\,\pi_L'\, \epsilon +2\, c_0^2 \,k^{-2}\, {\cal H}\,\frac{\pi_0'}{\bar\varphi'}\,\epsilon\,;
\end{split}
\ee
The following action describes the linear dynamics at the leading order of a slow-roll expansion in $\epsilon$:
\be
S_2 \equiv  
 \int dt \;d^3k\; \left( \ha {\pi'}^t \, {\cal K} \, \pi' +  {\pi'}^t \, {\cal D}_{\pi} \,
  \pi - \ha \pi^t\, {\cal M}_{\pi} \, \pi \right) \, , \quad \quad \pi = \begin{pmatrix}  \pi_L \\
  \pi_0 \end{pmatrix} \, ;
\label{slowact}
\ee
with
\bea
&& {\cal K} = \plm^2 \, \mathcal{H}^2 \, a^2 \, \epsilon \left(
\begin{array}{cc}
 4   \,  \left(1+c_1^2\right)\,k^2 & 0 \\
 0 & 8 \,  \,  c_0^2\;\bar{\varphi}
     '{}^{-2}
     \end{array} \right) \, ,
\\[.2cm]
&&
 {\cal D}_{\pi} =  -\frac{2 \,a^2 \,k^2  \,\plm^2  \,\mathcal{H}^2  \,\epsilon  \, \left(2  \,c_0^2 \,
   c_b^2+c_1^2\right)}{\bar\varphi '}\begin{pmatrix} 0 &1 \\
  - 1& 0 \end{pmatrix} \, , \\ [.2cm]
&&
{\cal M}_{\pi} = 4 \, \plm^2 \, a^2 \, \mathcal{H}^2 \,k^2\, \epsilon \left(
\begin{array}{cc}
 k^2 \left( -1+\frac{4}{3}\,c_2^2  -2  \,c_0^2 \,  c_b^4\right) & 
 -\frac{ 3\, \mathcal{H}\,  \left(c_1^2-2\,c_0^2\,c_b^2\right)\, \left(1+ \,c_b^2\right)}{2\;\bar\varphi '} \\
 -\frac{ 3\, \mathcal{H}\,  \left(c_1^2-2\,c_0^2\,c_b^2\right)\, \left(1+ \,c_b^2\right)}{2\;\bar\varphi '} & -\frac{    c_1^2}{\bar\varphi '{}^2} \\
\end{array}
\right) \,  .
\ea
Up to boundary terms, one can always take ${\cal D}$
antisymmetric. The peculiarity of (\ref{slowact}) is   the
  mixing of the two propagating DoF (degrees of freedom) present  at kinetic level due to the matrix ${\cal D}$  and at mass level being the matrix ${\cal M}$
non-diagonal. Such mixing is unavoidable unless the parameters
$c_b$ and  $c_2^2$, $c_1^2$  are unnaturally tuned, and  it is a key
property of a superfluid component in the solid at the origin of cross-correlations in the two and three points function of any scalar perturbation.
As a result,  the study of scalar linear perturbations is a  bit
involved and to get rid of the mixing by a suitable field redefinition few steps are needed. A similar system of coupled modes,
described by (\ref{slowact}), was encountered when studying the non-thermal production of gravitinos \cite{Nilles:2001fg}, multi-field
inflation \cite{Tolley_2010},
chromo-natural inflation \cite{Dimastrogiovanni_2013,
  Dimastrogiovanni:2012ew} and in effective theories of
inflation\cite{Bartolo:2015qvr}.  As far as we know, our analysis is the first complete one that does not rely on special choices of parameters. \\
The strategy to quantize the quadratic action will be the following. We
start from the original fields $\pi_{0,L}$ that describe physically
two Nambu-Goldstone modes around a non-trivial Lorentz breaking background solution. The quadratic action controlling the dynamics of such modes (\ref{slowact}) exhibits both kinetic (the presence of ${\cal D})$ and mass mixing effects (non-diagonal ${\cal M}$). A similar kinetic
mixing is also encountered in mechanical systems with gyroscopic
forces like the Coriolis force or in the presence of magnetic fields; it
is worth to stress that the ${\cal D}$ mixing can take place
when at least two fields are present. 
The first step is to make the fields canonical by a
time-dependent field redefinition $ \Pi={\cal K}^{1/2}\;\pi$ (\ref{canfield}).
At this level the corresponding Lagrangian $L(\Pi,\;\Pi')$  (\ref{lagrangian}) is characterized by non trivial ${\cal D}-mixing$ and a time-dependent non diagonal mass matrix.
The classical equations of motion correspond to a coupled system of second-order equations or, alternatively, to two decoupled fourth-order differential equations.
\\
 \begin{figure}[!tbp]
  \centering
  \includegraphics[width=.5 \textwidth]{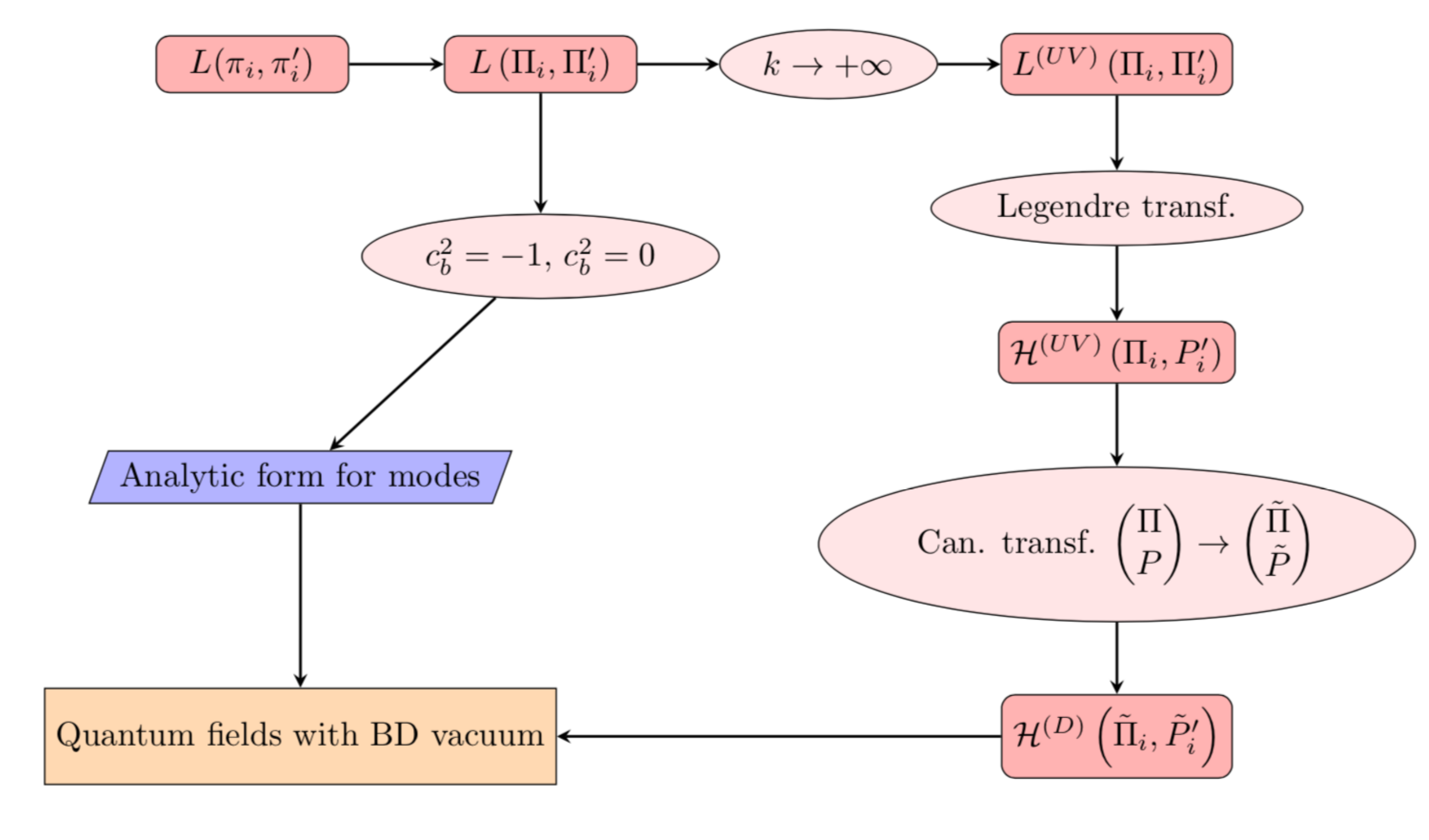}
  \caption{Schematic procedure for quantization}
  \label{pic:scheme}
  \end{figure}
The quantization of the system goes through the choice of the BD by  studying the Lagrangian in the UV  ($ k \to
\infty$) regime
(\ref{lagrangianUV}) where $k$ dominates over all other scales.
In this regime the mass term is diagonal and time independent.
Thanks to this feature we can write a decoupled system of quantum oscillators and quantize it with the usual canonical rules in Hamiltonian formalism ($L(\Pi,\;\Pi')\to L^{(UV)}(\Pi,\;\Pi')\to {\cal H}^{(UV)}(\Pi,\;P)$) (\ref{hamiltonian}). 
The quantum oscillator dynamics is recovered by a canonical transformation at Hamiltonian level 
(involving also the conjugate momenta) $  {\cal H}^{(UV)}(\Pi,\;P)\to
{\cal H}^{(UV)}_{diagonal}(\tilde\Pi,\;\tilde P)$
(\ref{newH}). Introducing the Hamiltonian formalism allows us to
decouple the two DoF  with a canonical transformation and to select
the BD vacuum.
The main steps are summarized in Figure \ref{pic:scheme}.  

\subsection{Quantization and Power Spectra} 
\label{quant-sect}

In order to define the Power Spectrum (PS) of a general
quantum scalar field $\xi(\textbf{x})$,  in Fourier space we set
\be
 \xi(\textbf{x})=\frac{1}{(2\,\pi)^3}\int\, d^3 k\, e^{i \, \textbf{k}\cdot \textbf{x}}\, \xi_{\textbf{k}}\,, \qquad \xi_{\textbf{k}}= \sum_{i=1}^2 \,\xi_{k}^{(cl)\,(i)}\, a_{\textbf{k}}{}^{(i)}+\xi_{k}^{(cl)\,(i)\,*}\, a_{-\textbf{k}}^{\dagger \,(i)} \,,  
\ee
where the (cl) subscript stands for classic solution, and the
latin index $i=1,\,2$ indicates the two indepedent scalar modes  whose  annihilation and creation operators obey the standard canonical commutation relations
\be
\left[a^{(i)}_{\textbf{k}},\, a^{\dagger}_{\textbf{p}}{}^{(j)}\right]=(2\,\pi)^3 \, \delta(\textbf{k}-\textbf{p})\,\delta^i_j \,. 
\ee 
Thus, the $\xi$  2-point function reads
\be
\langle \xi_\textbf{k}\, \xi_\textbf{p}\rangle = (2\,\pi)^3 \, P_{\xi}(k)\, \delta(\textbf{k}+\textbf{p})\,,  
\ee 
and the $\xi$ scale-invariant PS is defined by
\be
\begin{split}
{\cal P}_{\xi}= \frac{k^3}{(2\,\pi^2)}\,P_{\xi} & \equiv \frac{k^3}{(2\,\pi^2)}\left( |\xi_{k}^{(cl)\,(1)}|^2+|\xi_{k}^{(cl)\,(2)}|^2\right)\\
& = {\cal P}_{\xi}{}^{(1)}+{\cal P}_{\xi}{}^{(2)}\,.
\end{split}
\ee   
The first step to compute quantum correlators during inflation is to introduce the canonical field $\Pi$ defined by
\be
\Pi \equiv \begin{pmatrix} \Pi_{L} \\ \Pi_{0} \end{pmatrix} = {\cal
  K}^{1/2} \begin{pmatrix}  \pi_{L} \\ \pi_{0} \end{pmatrix} \, .
\label{canfield}
\ee
The conditions (\ref{fstab}) guarantee that the matrix ${\cal K}$ is
positive definite. Given that the matrix elements of ${\cal K}$ are
time-dependent, besides turning the kinetic term into a canonical one,
the transformation (\ref{canfield}) also alters the form of ${\cal D}$
and ${\cal M}$; thus, the quadratic Lagrangian in (\ref{slowact}),
when written in function of the new canonical fields becomes
\be
L_2 = \ha \Pi'{}^t \Pi' - \Pi^t \, {\cal D} \,
\Pi' - \ha \Pi^t\, {\cal M}\, \Pi  \; ;
\label{lagrangian}
\ee
where
\be
{\cal D}=k \,d\,
\begin{pmatrix}
0 & -1 \\
1 & 0
\end{pmatrix} \,, \qquad d= \frac{c_1^2+2 \, c_0^2 \;c_b^2}{2 \, \sqrt{2\, c_0^2 \,(1+c_1^2)}}\,;
\ee
\be
{\cal M}= \begin{pmatrix}
k^2\, \lambda_L{}^2 -6\, {\cal H}^2 & k\, {\cal H}\, \lambda \\
k\, {\cal H}\, \lambda & k^2\, \lambda_0{}^2-(1+3 \, c_b^2)(2+3\, c_b^2)\,
{\cal H}^2
\end{pmatrix} 
\,,
\label{mass2}
\ee
\be
\lambda_L^2= \frac{c_L^2-2 c_0^2 \,c_b^4}{1+c_1^2}\,, \qquad \lambda_0^2 = -\frac{c_1^2}{2\, c_0^2}\,, \qquad \lambda=\frac{ 2^{1/2}\left[c_0^2 \, c_b^2\, (1+3\, c_b^2)- c_1^2\right]}{ \sqrt{ c_0^2 \,(1+c_1^2)}}\,,
\ee
and we have defined
\be
\label{cl}
c_L^2 = -1 + \frac{4}{3} \, c_2^2 \, .
\ee
 At the leading order in slow-roll, the equations of motion
  are the following
\be
\label{coup_eq}
 \begin{split}
 & \Pi_L{}''-2 \, k \, d \Pi_0'+\Pi_0 \, k\, {\cal H}\, \lambda+\Pi_L\, \left(k^2 \, \lambda_L^2-6\, {\cal H}^2\right)=0\,,\\
 & \Pi_0{}''+2 \, k \, d \Pi_L'+\Pi_L \, k\, {\cal H}\,
 \lambda+\Pi_0\, \left(k^2 \, \lambda_0^2-(1+3\, c_b^2)\,(2+3 \, c_b^2)\, {\cal H}^2\right)=0\,.
 \end{split}
\ee
In order to quantize  (\ref{lagrangian}), 
 we need to remove the kinetic mixing introduced by
${\cal D}$. Our strategy is the following: in the UV, at very large $k$, ${\cal
  M}$ becomes diagonal and time-independent.
Thus, at very large $k$, the
original Lagrangian (\ref{lagrangian}) is equivalent to $L_2^{(\text{UV})}$, in accordance with the equivalence
principle. In that regime, by using a canonical transformation, one
can reduce the Hamiltonian associated to $L_2^{(\text{UV})}$ to a
system of two canonical free fields $\tilde{\Pi}=(\tilde{\Pi}_L, \, \tilde{\Pi}_0)$ linearly
related to $\Pi$
\be
L_2^{(\text{UV})}= \ha \tilde{\Pi}'{}^t  \tilde{\Pi}' -\ha \tilde{\Pi}^t \begin{pmatrix}  k^2 \,
  c_{s1}^2 & 0 \\
  0& k^2 \, c_{s2}^2  \end{pmatrix} \tilde{\Pi} \, .
\label{Losci}
\ee
Thus, the unique Fock space vacuum is the BD vacuum for the system. Details can be found in Appendix \ref{can-app}.
The existence of the BD vacuum requires  the frequencies
squared $\left \{ \omega^2_1= k^2 \, c_{s1}^2 \, , \;   \omega^2_2=
  k^2\, c_{s2}^2 \right \}$ to be strictly positive or equivalently
$c^2_{si} >0 \, , \; i=1,2$. In  addition we restrict ourselves to the
case of subluminal ``diagonal'' sound speeds:  $0< c_{si}^2<1$. The
conditions
(\ref{fstab}) are sufficient to have $c_{si}^2 >0$ and when
expressed in terms of (\ref{par}) gives~\footnote{We assume the null
  energy condition $1+w>0.$}
\be
  c_0^2>0, \, \qquad -1<c_1^2<0 \, , \qquad  c_L^2> 2 \, c_0^2 \,
  c_b^4   \, .
  \ee
We have checked that there is a large region of parameters $c_i$ where the
 stability conditions hold together with $c^2_{s1} \, , c^2_{s2}< 1$;
 moreover  in such a region, the first two conditions when rewritten in terms of $c_{s1}$ and $c_{s2}$,  become
  \be
  c^2_{s2}<  c_L^2 <c^2_{s1}  \, ,
  \label{relsp}
  \ee
where we choose the convention $c_{s2} < c_{s1}$. 
The equations of motion (\ref{coup_eq}) constitute a coupled 
system  of second order linear differential equations with time-dependent coefficients. Finding
explicit solution is not an easy task; of course, one could solve the
equations numerically. However, from a physical point of view, it is more transparent to
quantize the system focusing on the following values of $c_b$:
$c_b^2=-1$ and $c_b^2=0$   for which an analytic
  solution can be found. Neglecting SR corrections, the coupled system
  of second order equations can be written as two indepedent  fourth order equations for $\Pi_{0,L}$.\\
Remarkably, the case   $c_b^2=0$ and   $c_b^2=-1$ gives
the following identical equations  valid if
$c_1^2\neq0$
\be
\begin{split}
&\Pi_L{}^{(4)}+\left(c_{s1}^2+c_{s2}^2-\frac{12}{x^2}\right)\, \Pi_L{}''+\frac{24}{x^3}\, \Pi_L{}'+
\left(c_{s1}^2\;c_{s2}^2-\frac{6}{x^2}\,(c_{s1}^2+c_{s2}^2)\right)\, \Pi_L \,=0\,,\\
&\Pi_0{}^{(4)}+\left(c_{s1}^2+c_{s2}^2-\frac{4}{x^2}\right)\, \Pi_0{}''+\frac{8}{x^3}\, \Pi_0{}'+
\left(c_{s1}^2\,c_{s2}^2-\frac{2}{x^2}\,\left(c_{s1}^2+c_{s2}^2\right)-\frac{8}{x^4}\right)\, \Pi_0 \,=0\, ;
\end{split}
 \label{eq_mot}
 \ee
with $x=-k\, t$.
Note the presence in (\ref{eq_mot}) of the symmetry: $c_{s1}^2\leftrightarrow c_{s2}^2$. 
Analytic solutions are possible due to the absence of the terms $\Pi_i^{(3)}$ in the evolution equations.
The solutions can be
written as a linear combination of Bessel functions of order $5/2$ and
$3/2$; the integration constants are fixed by imposing that subhorizon,
 where $x   \gg1 $, the solution that  represents flat space modes matches the ones given in
 (\ref{piLexp}) and (\ref{pi0exp}); such a choice is equivalent to
 select the BD vacuum. Thus,  $\Pi_L$ and $\Pi_0$
   are quantum free (Gaussian)
 fields given by
 \be
 \begin{split}
&\Pi_{L\,k}=-i\,\sum_{j=1}^2 \,
a_{\textbf{k}}^{(j)}\,C_L^{(j)}\,\sqrt{\frac{\pi}{2}}\, c_{s\,
  j}{}^{\frac{1}{2}} \,\sqrt{-k\, t}\;
H^{(1)}_{\frac{5}{2}}\left(-c_{s\,j}\,k\,t\right)+\text{h.c.}\,, \\
&\Pi_{0\,k}=-\sum_{j=1}^2\,a_{\textbf{k}}^{(j)}\, C_0^{(j)}\,\sqrt{\frac{\pi}{2}}\, c_{s\, j}{}^{\frac{1}{2}}\, \sqrt{-k\, t}\; H^{(1)}_{\frac{3}{2}}\left(-c_{s\,j}\,k\,t\right)+\text{h.c.}\,;
\end{split}
\label{Piexp}
\ee
the expression for $\left \{ C_L^{(j)} \, , C_0^{(j)} \, ; j=1,2
\right \}$ can be found in Appendix \ref{can-app} and
$a_{\textbf{k}}^{(j)}$ are the creation operators for the fields
defined in eq.(\ref{diagonal_fields}).
In single field SR inflation, naturally two gauge invariant
scalar quantities can be considered when  studying the dynamics of
 superhorizon modes: the curvature $\zeta$ of the  $\rho=$
 const. hypersurface and the curvature ${\cal R}$ of hypersurface
 orthogonal to the scalar component of the fluid 3-velocity
 \be
 \begin{split}
  &\zeta = \Phi + {\cal H} \frac{\delta \rho}{\bar{\rho}'} \, ;\\
  & {\cal R} = \Phi + {\cal H} \,v .  
\end{split}
\ee
According to the Weinberg theorem \cite{Weinberg:2003sw}, in standard single field inflation, both $\zeta$ and $ {\cal R} $ are conserved and
coincides at superhorizon scales; as a result, the power spectrum of
primordial perturbations during inflation is given in terms of the
Fourier transform of the 2-point function of $\zeta$ or equivalently of  $ {\cal R} $. In our case, the Weinberg theorem does not hold and, besides the above curvature perturbations, additional gauge invariant scalar perturbations can be considered.
In particular the curvature $\zeta_n $ of constant particle number
$n$-hypersurface  ( keep in mind  that $n=n_\ell$), (\ref{defn})) and curvature of the $\varphi^0=$const.
hypersurface;
namely
\be
\zeta_n=-\Phi+\frac{{\cal
     H}}{\bar{n}'}\,\delta n \, , \qquad \qquad  
 {\cal R}_{\pi_0}= -\Phi+\frac{{\cal H}}{\bar\varphi'}\,\pi_0 \, .
\ee
The comoving curvature ${\cal R}_{\pi_0}$ is related to the superfluid component (\ref{svel}).
The expression of the various curvature perturbations in terms
$\pi_L$ and $\pi_0$ in a generic gauge can be found in Appendix \ref{gi-app}.
In the spatially-flat gauge (\ref{flatg}) we have that 
\be
\zeta_n=\frac{k^2}{3} \pi_L \, , \qquad  {\cal R} _{\pi_0} =
\frac{\cal H} {\bar{\varphi}'} \pi_0 \,.
\ee
The uniform curvature perturbation $\zeta$ can be obtained from
eq. (\ref{zeta}) at  leading order in SR 
\be\label{defZ}
 \zeta=\left(1-2\, c_0^2 \, c_b^2\right)\zeta_n-\frac{2}{3}\, c_0^2\,
 \frac{\pi_0'}{\bar\varphi'} ;
\ee
and similarly, from eq. (\ref{R}), for the comoving curvature
\be\label{defR}
{\cal R} =\frac{3 \, {\cal H}\,(1+c_1^2)}{k^2} \zeta_n'-c_1^2 \, {\cal
  R}_{\pi_0} \,.
\ee
As anticipated, by using (\ref{Piexp}), being ${\cal C}_{L/0}^{(j)}
\propto k^{-\frac{1}{2}}$, the power spectra of  $\zeta_n$, ${\cal
  R}_{\pi_0}$, $\zeta$ and ${\cal R}$ will be scale-free, up to
SR  corrections. Moreover, as shown in (\ref{R}) and (\ref{zeta}), $\zeta$ and ${\cal
  R}$ are linear combinations of $\zeta_n$ and ${\cal R}_{\pi_0}$ and
their time derivatives, the same conclusion applies to their spectral
indices. Thus, in the region $-1 \leq c_b^2 \leq 0$, all the relevant curvature perturbations have an almost scale-free PS. 
 From (\ref{Piexp}), (\ref{R}), (\ref{zeta}) and (\ref{super-rel}), at  leading order SR expansion,  we get for $c_b^2=-1\, , 0$
\bea\label{PZ}
 {\cal P}_{  \scriptscriptstyle \zeta_n} \!\!\!\!\!&=&\!\!\!\!\!
\frac{ {\cal P}_{SF}}{\left(c_b^2-c_L^2\right)^2 \left(
  c_{s1}^2-c_{s2}^2\right)} \left[\frac{\left(c_b^2-c_{s1}^2\right)^2 \left(
 c_L^2- c_{s2}^2  \right)}{c_{s1}^5 } +\frac{\left(c_b^2-c_{s2}^2\right)^2 \left(
 c_{s1}^2 -c_L^2 \right)}{c_{s2}^5 } \right]  ; \\
 {\cal P}_{\scriptscriptstyle{\cal R}_{\pi_0}}
\!\!\!\!\!&=&\!\!\!\!\! \frac{ {\cal P}_{SF}}{ 
  \left(c_{s1}^2-c_{s2}^2\right)}
\left[ \frac{\left(c_b^2-c_{s2}^2\right)^2}{c_{s1}  \left(c_L^2 -c_{s2}^2 \right)
 } +
\frac{\left(c_b^2-c_{s1}^2\right)^2}{c_{s2} 
  \left(c_{s1}^2-c_L^2\right) } \right]\,;
\ea
where we have introduced the scalar PS ${\cal P}_{SF}$ in canonical single field inflation
\be
{\cal P}_{SF} \equiv \frac{H_i^2}{8\, \pi^2\, M_{pl}^2 \,
  \epsilon} \, .
\ee
 where  $H_i$   denotes the value of the Hubble parameter during dS.
 For the cross-correlation we have
\be
\begin{split}
& {\cal P}_{{\scriptscriptstyle \zeta_n {\cal R}_{\pi_0}}} ={\cal
  P}_{SF} \, \frac{\left(c_{s1}^2-c_b^2\right)\left(c_{s2}^2-c_b^2\right)}{\left(c_L^2-c_b^2\right)\left(c_{s1}^2-c_{s2}^2\right)}
\left[ c_{s1}{}^{-3}  - c_{s2}{}^{-3} \right].
\end{split}
\ee
As we will discuss in section \ref{reheating}, for the simplest reheating scenario, the seed of primordial perturbations is given by the power spectrum of $\zeta_n$. 
Let us set
\be
{\cal P}_{\zeta_n}= {\cal P}_{SF}\, \bar{\cal P} \, ,
\ee
 where  $\bar{\cal P}$ is a dimensionless form factor
depending on $c_L^2$, $c_{s1}^2$, $c_{s2}^2$ and $c_b^2$ that  can be
read out from eq.(\ref{PZ}).   It is interesting to compare the above expressions with other existing models on the market.
General single field models, in the effective field theory
approach \cite{Cheung:2007st}, when the sound speed is different
from one, give  $\bar{\cal P}= c_s^{-1}$; while in adiabatic solid
inflation model $\bar{\cal P}$ reduces to $c_L^{-5}$ (see
(\ref{cl})). Thus $\bar{\cal P}$ can be interpreted as a sort of
{\it effective sound speed} parameter in order  to compare our
predictions with different inflationary models.
It should be stressed that the singular behavior of the PSs when $c_{s1}^2$ or
$c_{s2}^2$ is sent to zero or coincide, signals the simple fact that there is no
way to change the number of propagating DoF in a controlled way. This,
for instance, manifests trying to re-obtain the  PS for an adiabatic solid result from the supersolid  one by
imposing $c_{s\,i} \to c_L$,  leading to  divergence proportional to
$1/(c_{s1}^2-c_{s2}^2)$ as one can see from
(\ref{PZ}). 
\begin{figure}[!tbp]
  \centering
   \begin{minipage}[b]{0.42\textwidth}
    \includegraphics[width=\textwidth]{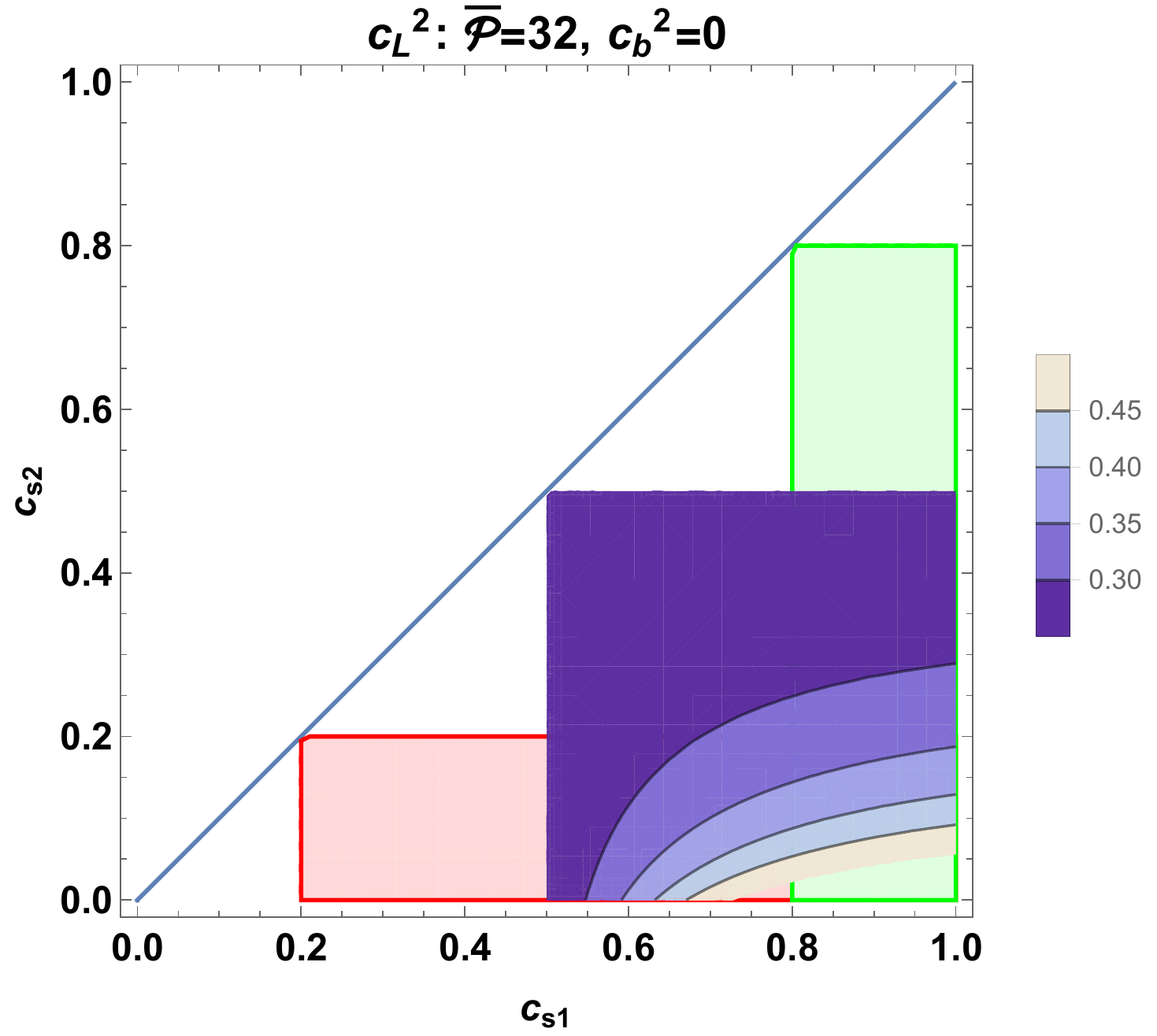}
    \subcaption{$c_b^2=-1$ }
  \end{minipage}
  \hspace{1cm}
  \begin{minipage}[b]{0.42\textwidth}
    \includegraphics[width=\textwidth]{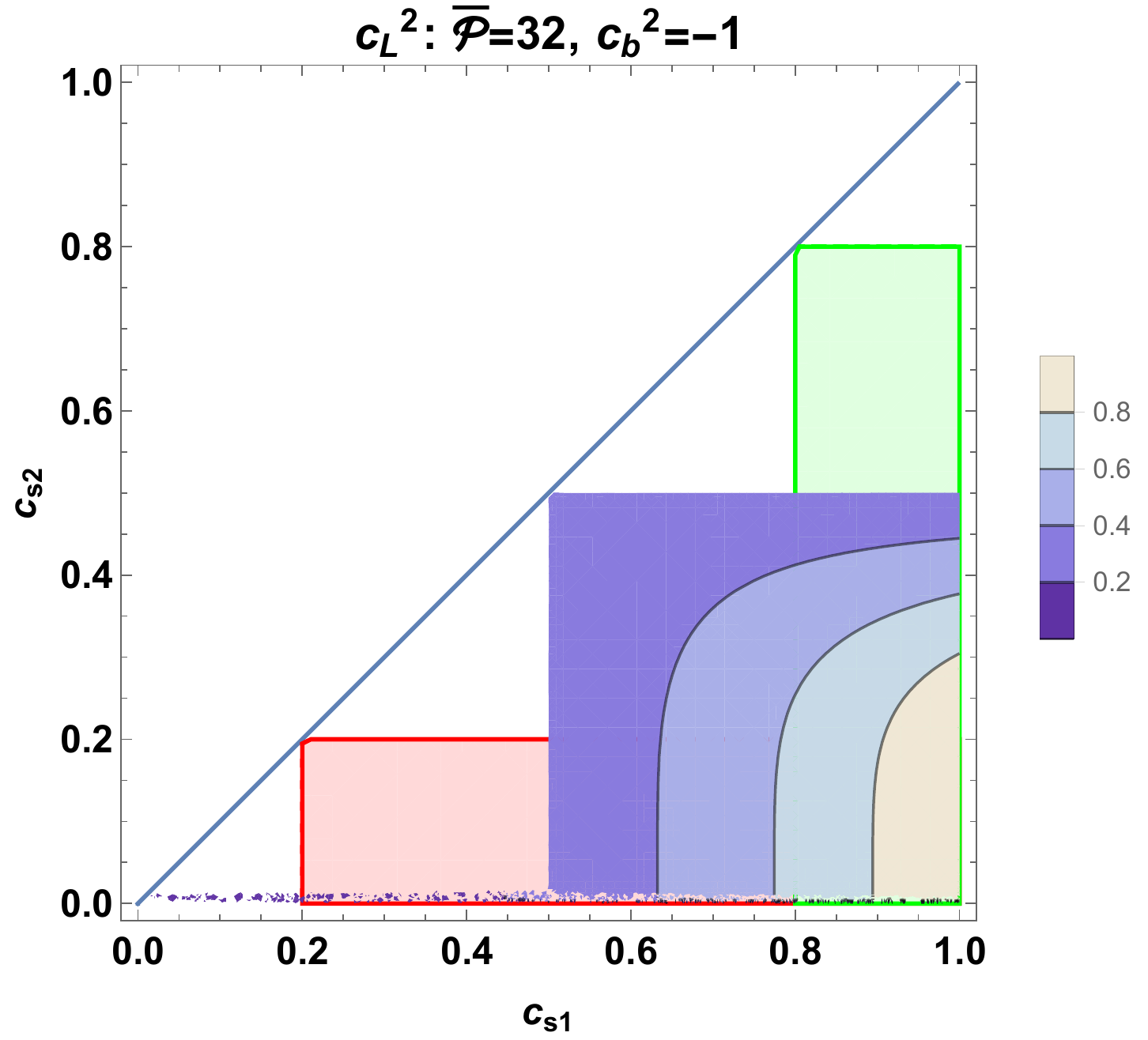}
    \subcaption{$c_b^2=0$ }
  \end{minipage}
\caption{$c_L^2$ as a function of $\left(c_{s1},\,c_{s2},\bar{\cal
      P}\right)$. The contour plot for $\bar{\cal P}=32$
  which maximizes the allowed region. The green curve represents the
  area covered for smaller values of $\bar{\cal P}$ while the red
  one shows  higher value of $\bar{\cal P}$.}
\label{cb01}
\end{figure}
\\
In the stability region, one can choose $\bar{\cal P}$ such that the amplitude of the $\zeta_n$ power spectrum is of order $10^{-9}$ as
required by observational constraints as  shown in Figure \ref{cb01}. We set $\bar{\cal P}$ to a constant, extracting $c_L^2$ as a function of $\left(c_{s2},\, c_{s1},\, c_b^2\right)$.
When one of the two diagonal sound speeds tends to the longitudinal one, for instance $c_{s1}\to c_L$, then $c_L$ reduces to its maximal value $\bar{\cal P}^{-\frac{1}{5}}$. The maximal allowed area corresponds to a maximal longitudinal speed $c_L=\frac{1}{2}$ and $\bar{\cal P}=32$.
Thus, by taking $5 < \bar{\cal P}<100$, there is a sufficient large
region in the parameters space spanned by $c_L$ and $c_{s1},\,c_{s2}$
to get a good agreement with data.

\begin{figure}[!tbp]
  \centering
   \begin{minipage}[b]{0.4\textwidth}
    \includegraphics[width=\textwidth]{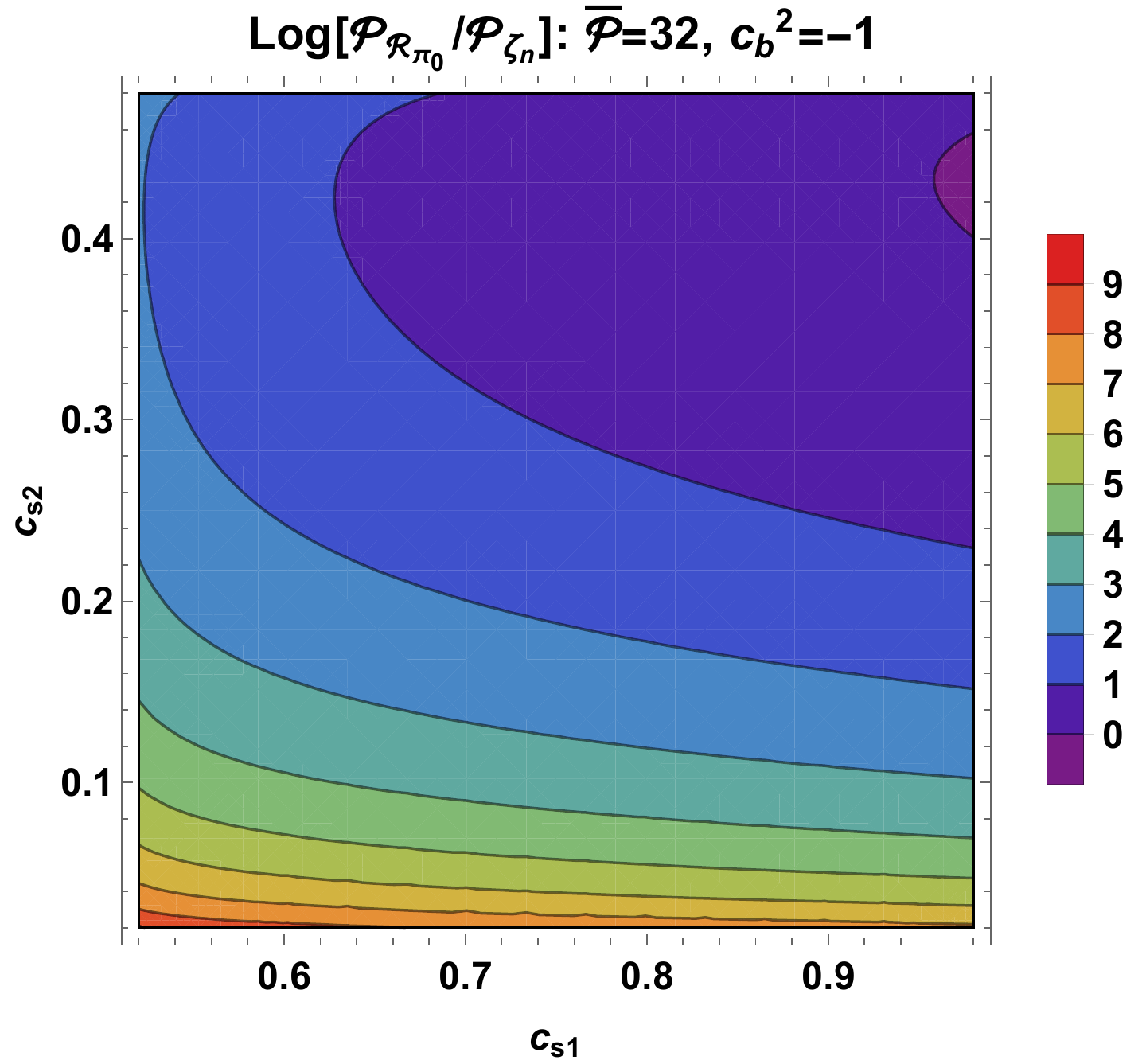}    
  \end{minipage}
  \hspace{1cm}
  \begin{minipage}[b]{0.41\textwidth}
    \includegraphics[width=\textwidth]{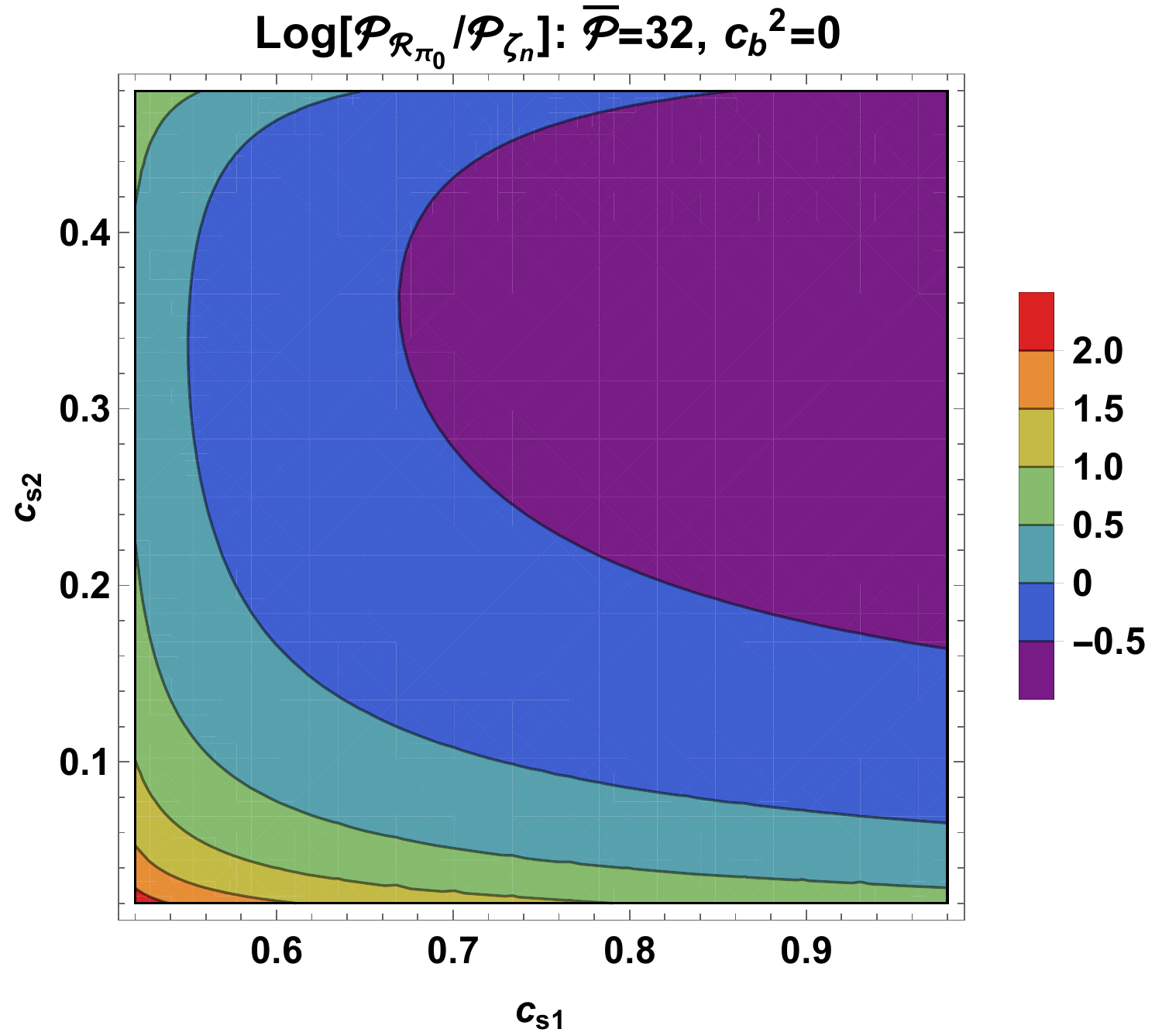}
  \end{minipage}
\caption{${\cal P}_{{\cal R}_{\pi_0}}/{\cal P}_{\zeta_n}$ plot. Note the ${\cal P}_{{\cal R}_{\pi_0}}$ enhancement particularly  pronounced in the $c_b^2=-1$ case.}
\label{P_ratio}
\end{figure}

It is useful to study the behavior of the various power spectra when
one of the sound speeds is much smaller
than the other: $c_{s2} \ll c_{s1}$ with $\bar{\cal P}$ fixed.
From our findings (\ref{PZ}),  we get
\be
\bar{\cal P}=\frac{ (c_{s1}^2 -c_{L}^2)}{(c_b^2-c_{L}^2)^2}\;\left(
\frac{c_b^4}{c_{s1}^2}\;\frac{1}{c_{s2}^5}+\frac{c_b^2\;(c_b^2-2\;c_{s1}^2)}{c_{s1}^4}\;\frac{1}{c_{s2}^3}+
\frac{(c_b^2-c_{s1}^2)^2}{c_{s1}^6}\;\frac{1}{c_{s2} } 
\right)+{\cal O}(c_{s2}^0) \, .
\ee
This gives, for the different values of $c_b$ that we are using, the approximate relations
\bea\label{cLs2}
&&c_L^2\simeq c_{s1}^2+ (1-\bar{\cal
  P}\,c_{s1}^5)\;\frac{(1+c_{s1}^2)^2}{c_{s1}^3}\,c_{s2}^5+{\cal
  O}(c_{s2}^6)\qquad {\rm for}\qquad c_b^2=-1 \, ;
\\
&&c_L^2\simeq c_{s1}^2+(1-\bar{\cal P}\,c_{s1}^5)\,c_{s1}\;c_{s2}+{\cal O}(c_{s2}^2)
\qquad {\rm for}\qquad c_b^2=0 \, .
\ea
Self consistency requires  $c_{s1}^2\geq c_L^2$ that implies
$\bar{\cal P}\geq1/c_{s1}^5$; thus if we take  $\bar{\cal P} =32$,
the two speeds of sound  are in the region: $0\leq c_{s2}\ll  1/2\leq
c_{s1}\leq1$.
It is precisely the constraint on   $\bar{\cal P}$ that introduces a
dramatic asymmetry, boosting ${\cal P}_{\scriptscriptstyle  {\cal
    R}_{\pi_0}}$. For  $c_b^2=-1$, ${\cal P}_{{\cal R}_{\pi_0}}$ is naively enhanced by a factor $1/c_{s2}$
\be
{\cal P}_{{\cal R}_{\pi_0}}= {\cal P}_{SF}\;\frac{(1+c_{s1}^2)^2}{c_{s1}^2\;(c_{L}^2-c_{s1}^2)\;c_{s2}}+{\cal
    O}(c_{s2}^0) \, ;
  \ee
however taking into account (\ref{cLs2}), an extra enhancing factor $1/c_{s2}^6$ is introduced; namely
\be{\cal P}_{{\cal R}_{\pi_0}}= {\cal P}_{SF}\;\frac{c_{s1}}{\left( c_{s1}{}^5\,\bar {\cal P}-1 \right) } \,\frac{1}{c_{s2}{}^6}+{\cal O}(c_{s2}^{-5}) \, .
\ee
Similarly, for $c_b^2=0$,  an enhancement  from $1/c_{s2}$ to
$1/c_{s2}^2$ is obtained
 \bea
 {\cal P}_{\scriptscriptstyle   {\cal R}_{\pi_0}}= {\cal P}_{SF}\;\frac{c_{s1}^2}{c_{s1}^2-c_L^2}\;\frac{1}{c_{s2}}+{\cal O}(c_{s2})
   \to  {\cal P}_{SF}\;\frac{c_{s1}}{\bar{\cal
        P}\;c_{s1}^5-1}\;\frac{1}{c_{s2}^2}+{\cal O}(c_{s2}^{-1}) \, .
\ea
So, the constraint imposed by the observed $ {\cal P}_{\zeta_{n}} $ increases the sensitivity 
to small sound speeds of  ${\cal P}_{\scriptscriptstyle {\cal R}_{\pi_0}} $ \footnote{ For the cross-correlation ${ \cal
     P}_{{\scriptscriptstyle \zeta_n {\cal R}_{\pi_0}}} $ we find the following expansion
 \be 
{\cal P}_{ \scriptscriptstyle  \zeta_n {\cal R}_{\pi_0}}= \frac{ H_i^2}{8\, \pi^2
  \,M_{pl}^2 \,\epsilon }
\begin{cases}    
      \frac{1}{c_L^2 \, c_{s2}} +O(c_{s2}^{0}) & c_b^2 =0\\[.4cm]
  \frac{1}{c_{s1}{}^2 \, c_{s2}{}^3} +O(c_{s2}^{-2}) &  c_b^2 =-1
\end{cases}
\ee
}.
Thus, at fixed ${\cal P}_{\zeta_n}$,
 the ${\cal P}_{{\cal R}_{\pi_0}}$   becomes dominant being proportional to negative powers of $c_{s2}$, see Figure \ref{P_ratio}. 
Let us briefly recap the relevant used parameters.\\
The quadratic Lagrangian contains  5 mass parameters (\ref{quadflat},
\ref{par})   or equivalently  $c_{0,1,2,3,b}^2$ ; it is
convenient to replace $c_4^2$ by $c_b^2$ in (\ref{cb}).
 On a  dS background $c_s^2=-1$, thus (\ref{dsss}) fixes $c_3^2$ to be
\begin{equation*}
 c_3^2=\frac{1}{2}+c_0^2\;c_b^4+\frac{c_2^2}{3}\,.
\end{equation*}
 In order to generate flat power
 spectra in a slow-roll regime requires that $-1\leq c_b^2\leq 0$.
 Moreover, we were able to find an analytic solution for the modes at
 any time $t$ only for $c_b^2=-1  $ and $c_b^2=0$; such values will be
 considered in the rest of the paper. \\ It is convenient to trade the three independent  parameters $c_{0,1,2}^2$ 
to $c_L^2$, $c_{s1}^2$ and $c_{s2}^2$. 
\\
While
$c_L^2=-1+\frac{4}{3}\,c_2^2\,$
can be interpreted as the longitudinal speed typical of Solid Inflation, the other two are the sound speeds corresponding to the two DoFs without any mixing (basically harmonic oscillators) described by (\ref{Losci}). 
Finally, by matching the amplitude of the scalar power spectrum to the observed value of $10^{-9}$ one can fix $c_L^2$ as a function of the remaining two free parameters $c_{s1}^2$ and $c_{s2}^2$.
\begin{table}[htp]
\begin{center}
\begin{tabular}{|c|c|c|c|}
\hline
Constraints & existence of dS   & Analytic modes \& $n_s \sim 1$ & PS amplitude\\
\hline
                  & $c_3^2= \frac{1}{2}+c_0^2\;c_b^4+\frac{c_2^2}{3}$ &
                                                           $c_b^2=-1,\,c_b^2=0$
                                                                 &
                                                                   ${\cal
                                                                   P}_{\scriptscriptstyle
                                                                   \zeta_n}(c_{s 1 }^2 ,\,c_{s 2 }^2,\,c_L^2)=10^{-9}$ \\
\hline
\hline
Free parameters& $c_{0,1,2, b}^2$ & $c_{0,1,2 }^2 $ & $c_{s 1 }^2 ,\;c_{s 2 }^2 $\\
  $c_{0,1,2,3,4}^2$ & & &  \\
 \hline
\end{tabular}
\end{center}
\label{freeconst}
\caption{Independent parameters and constraints for supersolids.
}
\end{table}
Let us note that our results, when comparable, do not agree
with the one in~\cite{Bartolo:2015qvr}. The reason is the
missing parameter $c_b$ (see footnote 3) and the treatment of the extra scalar degree of freedom in addition to the one present in solid inflation~\cite{Endlich:2012pz}.\\
\begin{figure}[!tbp]
  \centering
   \begin{minipage}[b]{0.42\textwidth}
    \includegraphics[width=\textwidth]{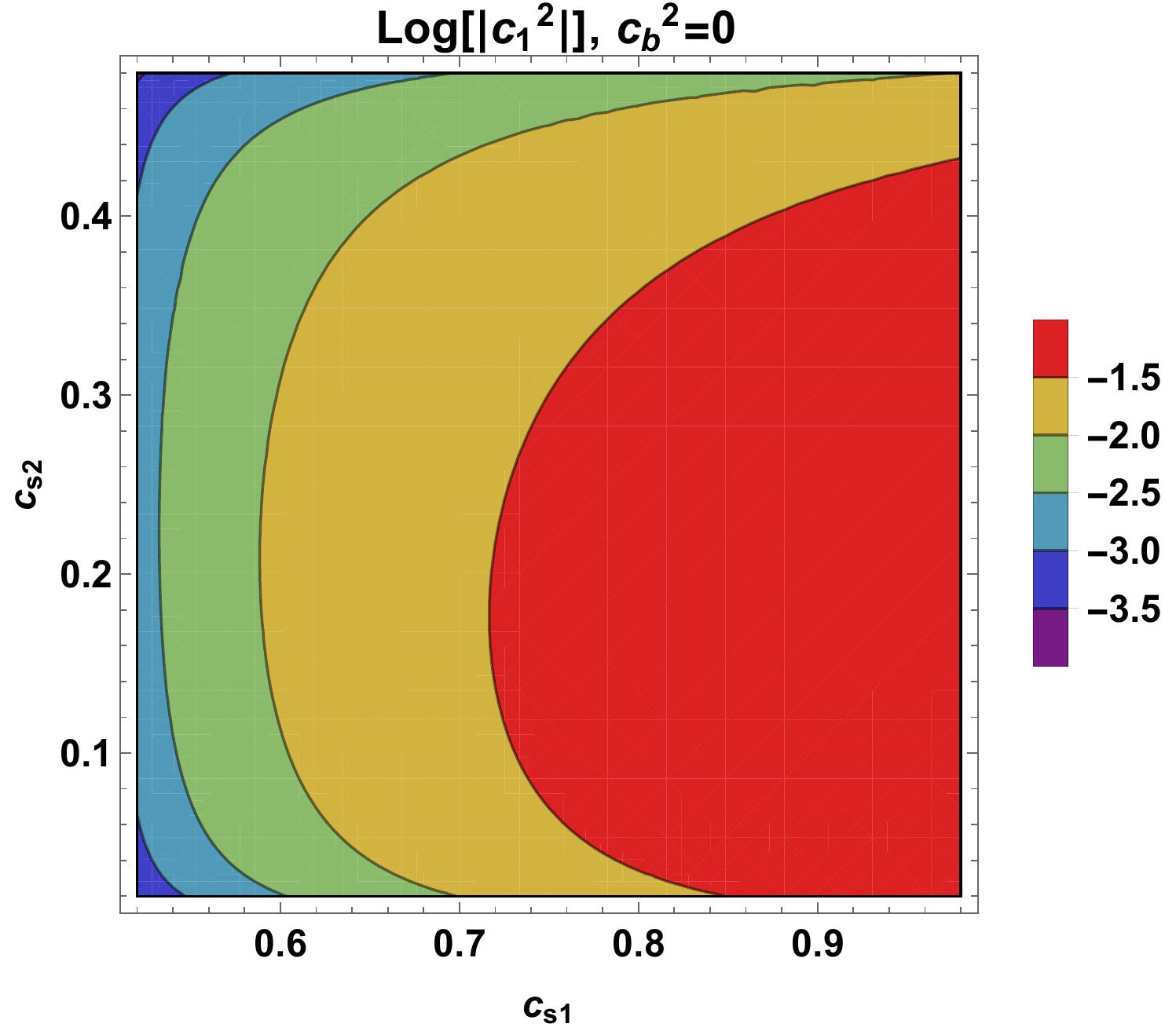}
  \end{minipage}
  \hspace{1cm}
  \begin{minipage}[b]{0.42\textwidth}
    \includegraphics[width=\textwidth]{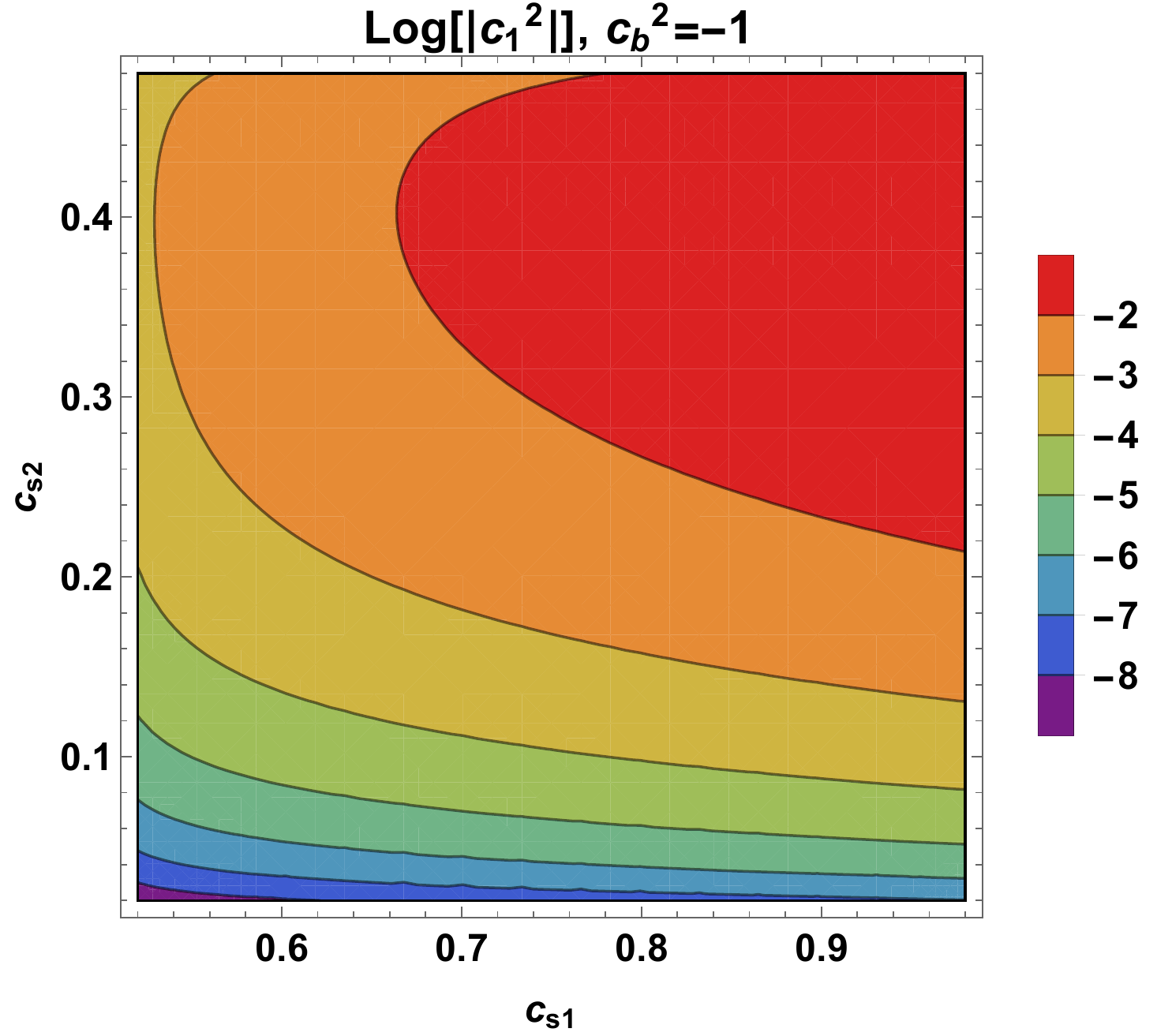}
  \end{minipage}
\caption{$c_1^2$ logarithmic contour plot for $c_b^2=-1$ and $c_b^2=0$. The parameter $\bar{\cal P}$ is set to $32$.}
\label{c1}
\end{figure}
\begin{figure}[!tbp]
  \centering
   \begin{minipage}[b]{0.42\textwidth}
    \includegraphics[width=\textwidth]{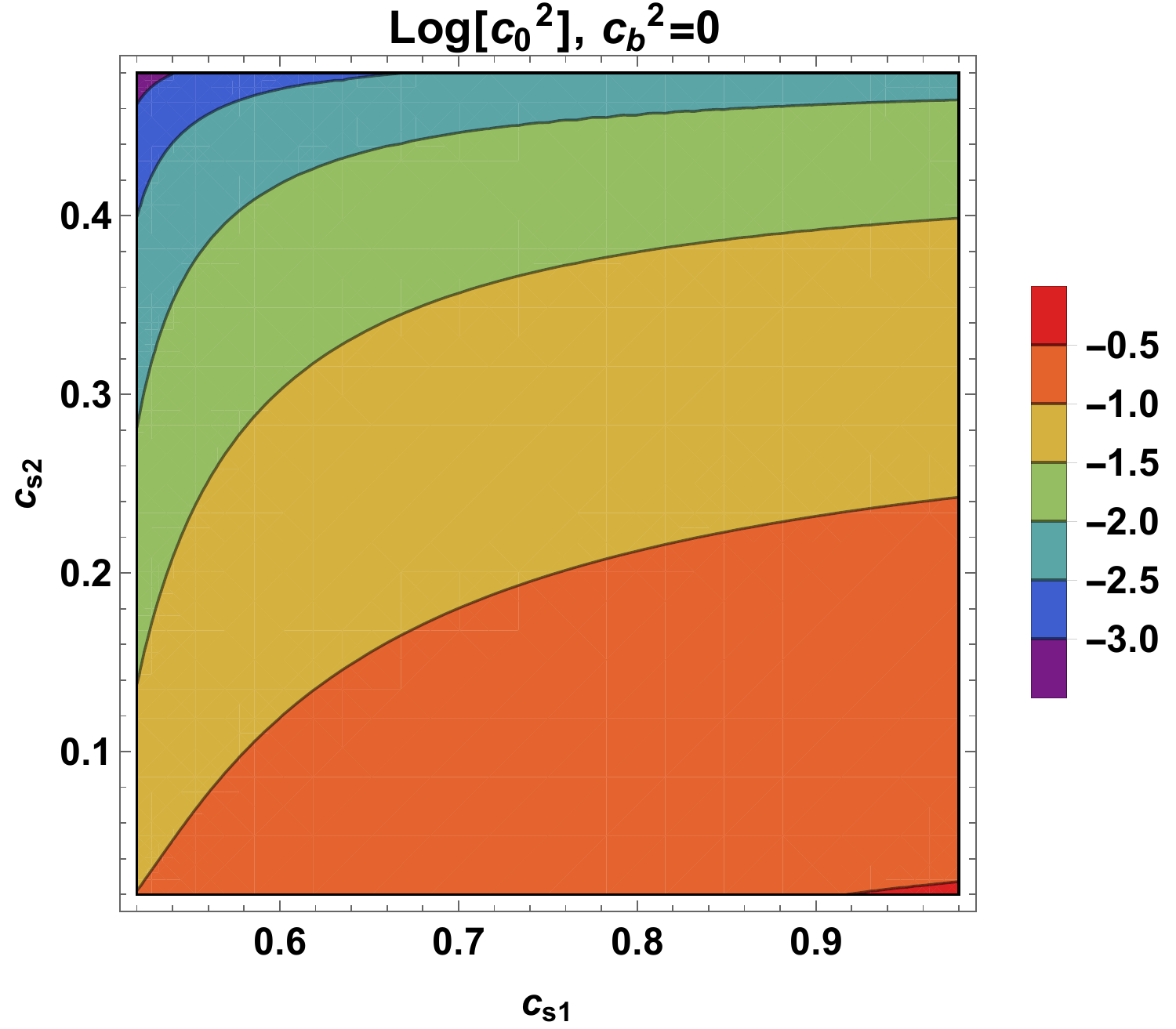}
  \end{minipage}
  \hspace{1cm}
  \begin{minipage}[b]{0.42\textwidth}
    \includegraphics[width=\textwidth]{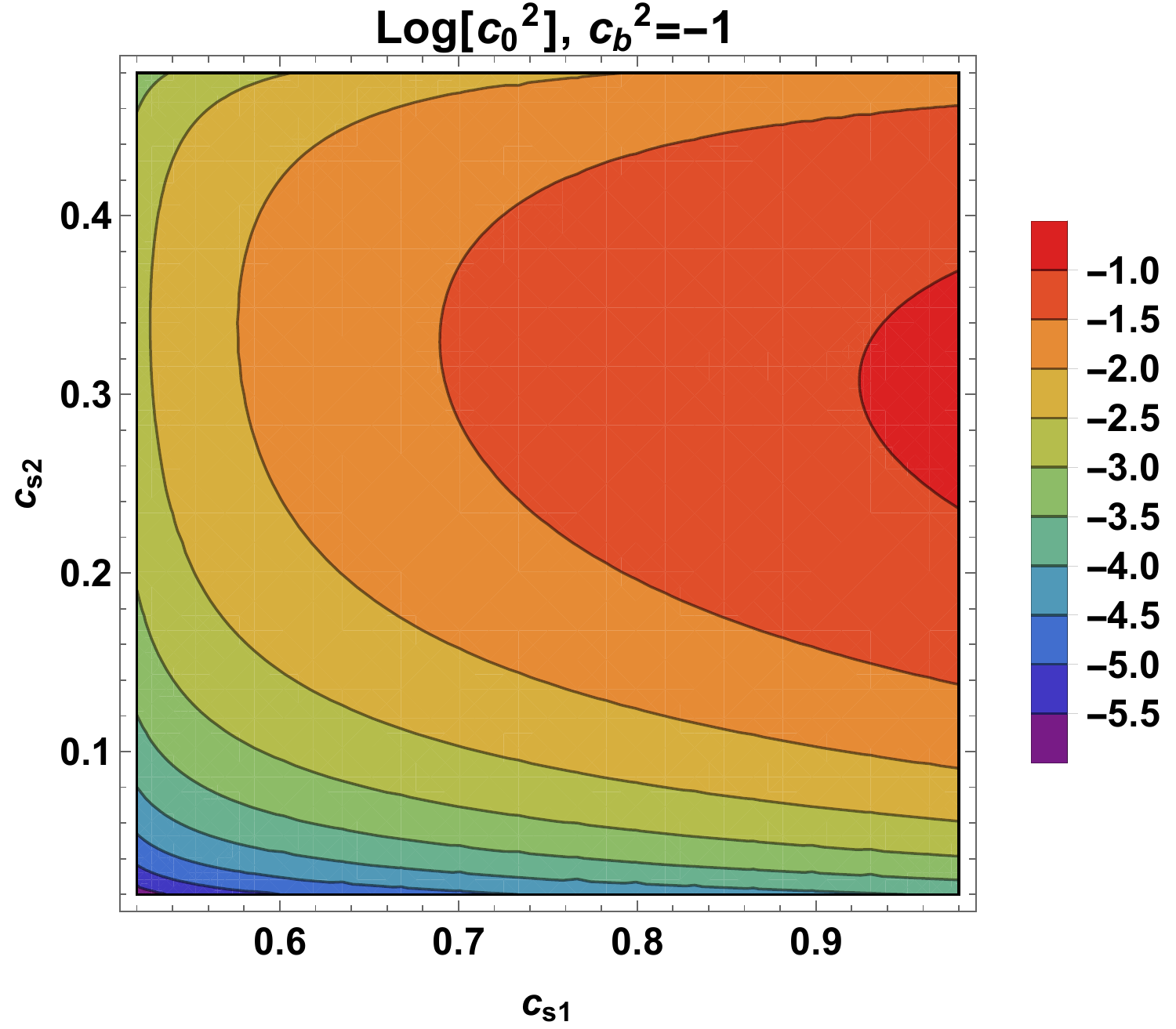}
  \end{minipage}
\caption{$c_0^2$ logarithmic contour plot for $c_b^2=-1$ and $c_b^2=0$. The parameter $\bar{\cal P}$ is set to $32$.}
\label{c0}
\end{figure}
Taking   $c_{s1}$ and $c_{s2}$ as independent parameters, the
 behavior of  $c_{1}$ and $c_0$ is important in the study of the amount of
 isocurvature perturbations and the secondary gravitational waves
 production. The complete expression is given in Appendix
 \ref{can-app}, we quote here the leading contribution for small $c_{s2}$ 
 \be 
\label{c_01_as}
\begin{split}
&c_{1}^2= \begin{cases}  \left(\frac{1}{c_{s1}}-\bar{\cal P} \,
    c_{s1}^4\right)\,c_{s2}^5+O(c_{s2}^7) 
    \\[.2cm]
 c_{s1}^3\frac{(1-\bar{\cal P}\,
   c_{s1}^5)}{(1+c_{s1}^2)^2}\,c_{s2}+O(c_{s2})
\end{cases} \, \quad
c_0^2= \begin{cases} 
   -\frac{1}{2}\,\left(\frac{1}{c_{s1}}-\bar{\cal P} \,
    c_{s1}^4\right)\,c_{s2}^3+O(c_{s2}^5) &\;\;\;\; c_b^2=-1 \\[.2cm]
   \frac{1}{2}\, c_{s1}^2+O(c_{s1}) & \;\;\;\;c_b^2=0
\end{cases} \, .
\end{split}
\ee
Note that we have used that ${\cal P}_{\zeta_n} =\text{const.}\sim
10^{-9}$ as a fixed constant value.
Thus, both $c_0$ and $c_1$ are suppressed for small $c_{s\,2}$
 as shown  in figures \ref{c1} and \ref{c0}.

\subsection{Slow-roll Corrections at Superhorizon Scales}
\label{sec:slowcorr}

In this section, we give the slow-roll corrections of the primordial PS,  focusing on the two exact solutions obtained for $c_b^2=-1$ and $c_b^2=0$.
Being the superhorizon behavior determined by the $c_b^2$ value even
at the leading order, the analysis of the large scales $\Pi_L$ and
$\Pi_0$ fields in dS approximation is fundamental in order to get
the parameter space where powerspectra are scale-free.
By manipulating the system of second order coupled equations
(\ref{eq_mot}), in the large scale limit $x \ll 1$ , we can get the following two independent fourth-order equations:
\bea
&&\Pi_L^{(IV)}+\frac{2}{x}\,\Pi_L^{'''}-\frac{1}{x^2}\left[8+9\, c_b^2\, (1+c_b^2)\right]\,\Pi_L^{''}+\frac{12}{x^3}\,\Pi_L^{'}+\frac{54 \, c_b^2 \, (1+c_b^2)}{x^4}\Pi_L=0\,,
 \\\nonumber
&&\Pi_0^{(IV)}+\frac{2}{x}\,\Pi_0^{'''}-\frac{1}{x^2}\left[8+9\, c_b^2\, (1+c_b^2)\right]\,\Pi_0^{''}+
\left[2+9\, c_b^2 \, (1+c_b^2)\right]\,\left[\frac{2}{x^3}\,\Pi_0^{'}
+\frac{4}{x^4}
\,\Pi_0\right]=0\,.
\ea 
The solutions can be expressed in the following form
\bea
\label{leading_tilt}
 \Pi_L = \frac{1}{\sqrt{k}}\; \left({\cal C}_{L,1} \;x^{-2 } +
 {\cal C}_{L,2} \;x^{3} +
 {\cal C}_{L,3} \;x^{3\;(1+c_b^2)} +
 {\cal C}_{L,4} \;x^{-3\;c_b^2}\right)\\
  \Pi_0 = \frac{1}{\sqrt{k}}\; \left( {\cal C}_{0,1}\; x^{2+3\;c_b^2} +
 {\cal C}_{0,2} \;x^{-1-3\;c_b^2} +
 {\cal C}_{0,3}\; x^{-1} +
 {\cal C}_{0,4}\; x^{4}\right)
\ea
where the ${\cal C}$ coefficients are unspecified constant at this stage. 
The ${\cal C}_{L/0,1/2}$ refer to homogeneous solutions while ${\cal
  C}_{L/0\,3/4}$ to particular solutions of the original system
(\ref{eq_mot}).
Thus, we can understand the effect of $c_b^2$  on the superhorizon
evolution by obtaining particular solutions for  $\Pi_L$, sourced by $\Pi_0$  and vice versa.
\begin{table}[htp]
\begin{center}
\begin{tabular}{|c||c|c| c|c|c|}
\hline
         &  $c_b^2< -\frac{5}{3}$  &  $-\frac{5}{3}< c_b^2 < -1$   &$-1< c_b^2 < 0$ &$0< c_b^2< \frac{2}{3}$ & $c_b^2 < \frac{2}{3}$ \\\hline\hline
$\Pi_L$& ${\cal C}_{L,3} \;\frac{x^{3(1+c_b^2)}}{\sqrt{k}} $    &\multicolumn{3} { | c | }{ ${\cal C}_{L,1}\;\frac{x^{-2}}{\sqrt{k}}   $ } &${\cal C}_{L,4}\;\frac{x^{-3\,c_b^2}}{\sqrt{k}}   $ \\\hline
$\zeta_n$& ${\cal C}_{L,3}\;\frac{x^{(5+3\,c_b^2)}}{k^{3/2}}   $
                                   &\multicolumn{3} { | c | }{ ${\cal
                                     C}_{L,1} \;\frac{1}{k^{3/2}} $
                                     Flat PS} &${\cal C}_{L,4}  \;\frac{x^{(2-3\,c_b^2)}}{k^{3/2}}$ \\\hline\hline
$\Pi_0$&\multicolumn{2} { | c | }{ ${\cal C}_{0,1}  \;\frac{x^{(2+3\,c_b^2)}}{\sqrt{k}} $}   &${\cal C}_{0,3}  \;\frac{x^{-1}}{\sqrt{k}} $  &\multicolumn{2} { | c | }{${\cal C}_{0,2}   \;\frac{x^{-(1+3\,c_b^2)}}{\sqrt{k}}$} \\\hline
${\cal R}_{\pi_0}$&\multicolumn{2} { | c | }{ ${\cal C}_{0,1}
                    \;\frac{x^{3(1+c_b^2)}}{k^{3/2}} $}   &${\cal
                                                            C}_{0,3}
                                                            \;\frac{1}{k^{3/2}}
                                                            $
                                                            Flat PS &\multicolumn{2} { | c | }{${\cal C}_{0,2} \;\frac{x^{-3\,c_b^2}}{k^{3/2}} $} \\\hline
\end{tabular}
\end{center}
\caption{Leading $t$ and $k$ behaviour for $x \to 0$ of
  $\Pi_{L,0}$ and  $\zeta_n$, ${\cal R}_{\pi_0}$ for different  values of  $c_b^2$.}
\label{T0L}
\end{table}%

\begin{table}[htp]
\begin{center}
\begin{tabular}{|c||c|c| c|c|c|}
\hline
       & $c_b^2 = -1$   &$-1< c_b^2 < 0$ &$  c_b^2=0$   \\\hline\hline
$\zeta_n$& 
 \multicolumn{3} { | c | }{ ${\cal C}_{L,1} \; k^{-3/2}  $}   \\\hline\hline
${\cal R}_{\pi_0}$&  
$({\cal C}_{0,1}+{\cal C}_{0,3}) \; k^{-3/2}  $  &
${\cal C}_{0,3} \; k^{-3/2}  $   &
 $({\cal C}_{0,3}+{\cal C}_{0,2} )\;k^{-3/2} $ \\\hline
\end{tabular}
\end{center}
\caption{Inside the region of flat PS $(-1\leq c_b^2\leq0)$ we give the late time structure of 
 $\zeta_n$ and ${\cal R}_{\pi_0}$. }
\label{TZR}
\end{table}%
The boundary values for
$c_b^2=-\frac{5}{3},\,-1,\,0,\,\frac{2}{3}$ are shown in  Table \ref{TZR}.
The relation between $\zeta_n$ and  ${\cal R}_{\pi_0}$  to the canonical fields asymptotically implies the following $k$ and time dependence
\be
\begin{split}
&\pi_L\propto \frac{t^2}{k}\;\Pi_L \quad \to\quad  \zeta_n\propto k^2\;\pi_L\propto k\;t^2\;\Pi_L
\\
&\pi_0\propto t^{1+3\;c_b^2}\;\Pi_0 \quad \to\quad {\cal R}_{\pi_0}\propto t^{-3\;c_b^2}\;\pi_0\propto t \;\Pi_0
\end{split}
\ee
Taking into account the various transformations, the almost scale-free PS
of $\zeta_n$ and ${\cal R}_{\pi_0}$  is obtained when the coefficient ${\cal C}_{L,1}$  and ${\cal C}_{0,3}$ dominate.
Thus, at the leading order, one realizes that  $\zeta_n$
is also always almost   scale-free in the region $-\frac{5}{3}
\le c_b^2 \le \frac{2}{3}$, while
 ${\cal R}_{\pi_0}$  selects a smaller region $-1 \le c_b^2
\le 0$. We will focus on this last region. 
Recall that ${\cal R}$ and $\zeta$  are simple  functions of $(\zeta_n\ ,\;{\cal R}_{\pi_0})$ and their time derivative, see  (\ref{defZ}, \ref{defR}).

Finally, let us outline the form of the relevant  scalar fields at
super-horizon scales, obtained by computing the next to leading
slow-roll corrections of the canonical normalized fields:
\be
\begin{split}
\label{structure}
\zeta_n &={\cal A}_{\zeta_n}{}^{(ad)} \,(-H_i\,t)^{\alpha^{(ad)}}\, k^{\frac{1}{2}\left[(n_{s }^{(ad)}-1)-3\right]}\,\\ 
{\cal R}_{\pi_0},\, {\cal R},\, \zeta &= 
{\cal A}_{X}^{(ad)} \,(-H_i\,t)^{\alpha^{(ad)}}\, 
k^{\frac{1}{2}\left[(n_{s }^{(ad)}-1)-3\right]}+{\cal A}_{X}^{(en)} \,(-H_i\,t)^{\alpha^{(en)}}\, k^{\frac{1}{2}
\left[(n_{s }^{(en)}-1)-3\right]} \, ;
\end{split}
\ee
the explicit form of the constants ${\cal A}_a$ are not relevant here. \\
Let us explain what superscripts ($ad$) and ($en$) mean. The ($ad$) part of a field stands for its adiabatic part,  in the sense that its adiabatic-tilt $n_{s}^{(ad)}$ will not be affected by the presence of the $c_b^2$ parameter. This parameter is strictly related to the presence of propagating superfluid density per lattice site (i.e. non-barotropic) perturbations. 
On the contrary, the superscript ($en$) stands for the entropic part of a field, and the related tilt $n_{s}^{(en)}$ will be $c_b^2$ dependent.
A crucial feature is that the behavior of $\zeta_n$ on superhorizon scales is determined by a single purely adiabatic power law
given in terms of $\alpha^{(ad)}$ and $n_s^{(ad)}-1$.  As we will show in the next section,  in the case of an instantaneous reheating, it is precisely $\zeta_n$ that determines the transition to the  radiation era, setting the adiabatic part of the initial conditions; moreover, the non-adiabatic part will be determined by the difference $\zeta-\zeta_n$. If $c_b^2$ is far from the interval $[-1, \,0]$, the time dependence from $\alpha^{(en)}$ will overwhelm the homogeneous ${\cal R}_{\pi_0}$ solutions, leaving only its particular adiabatic solution. In practice, when we are far
enough from the boundary values of $c_b^2$, also the other fields will
be single-tilted; however in this region, the link between super and subhorizon amplitudes needs to be computed numerically.\\ 
On the contrary, at the leading order in slow-roll, when $c_b^2=0,
-1$, we get analytical solutions in the form of almost scale-free
power laws for all relevant scalars
\be
{\cal R}_{\pi_0},\, {\cal R},\, \zeta\; \to\; \left({\cal A}_{X}^{(ad)}+{\cal A}_{X}^{(en)}\right) \, k^{-\frac{3}{2}}\,.   
\ee 
The above form was used in the previous section to compute the leading order amplitudes of the primordial PS for ${\cal R}_{\pi_0}$. In practice, when $c_b^2=0,-1$ we cannot discriminate between the adiabatic and entropic ${\cal R}_{\pi_0}$  parts.
Such a degeneracy is removed by next to leading slow-roll corrections. In the case of an almost instantaneous reheating, the slow-roll leading order computation of the $\Gamma_{\Lambda \text{CDM}}$, primordial
Non-Gaussianities and GWs back-reaction will be sufficient.\\
For completeness, we give the result of next to leading  slow-roll
corrections
\be
\begin{split}
&\qquad\qquad n_{s}^{(ad)}=1+ 2\, \epsilon_i \, c_L^2 -\eta \,
 , \qquad \alpha^{(ad)}=\frac{4}{3}\, c_T^2\,\epsilon,
 \\
&  \begin{cases} 
&   n_{s}^{( en)}=7+6\, c_b^2 \,(1+ \epsilon_i)+\eta \,, \;\; \alpha^{(en)}=3+3\, c_b^2\,(1+\epsilon)+\epsilon+\eta\,,
\qquad \text{dominant if} \;c_b^2 \approx -1\,, \\
   & n_{s}^{( en)}= 1-6\, c_b^2 \,(1+ \epsilon_i)-\eta\,, \;\; \alpha^{(en)}=-6\,c_b^2\,(1+\epsilon)-\eta\,,
   \qquad \text{dominant if} \;c_b^2 \approx 0\,.
\end{cases} 
\end{split} 
\ee
The ($ad$) tilt is formally the same as the one found in \cite{Endlich:2012pz}, being obtained by solving the same {\it superhorizon} equations of motion. However, starting from a supersolid, the solid inflation limit does not exist: the extra degree of freedom cannot be smoothly switched off. 
In principle, two ($en$) tilts are present in the other fields and exists only on the edges of the region.\\
Let us sketch the main steps to get the above slow-roll corrections on superhorizon scales:
\begin{enumerate}
\item Trade the system of coupled equations for $\left(\zeta_n,\,
    {\cal R}_{\pi_0}\right)$ for an equivalent but simpler to analyze
  involving $\left({\cal R},\, \delta \sigma\right)$;
\item Find the canonical fields $({\cal R}_c,\, \delta \sigma_c)$ and
  find the leading superhorizon behaviour at the leading order in
  slow-roll as  done for $\left(\Pi_L,\, \Pi_0\right)$. Define the
  ($ad$) part of ${\cal R}$ as its ``homogeneous'' (e.g. $c_b$ independent)
  component (coherent with the fact that $\zeta_n$ is a purely ``adiabatic'' field) and the ($en$) part of ${\cal R}$ as the $\delta\sigma$-sourced solutions. 
\item Compute the ${\cal R}$ and $\delta \sigma$ slow-roll corrections on superhorizon scales.
\item Degeneracy breaking: 
\begin{itemize}
\item for $c_b^2=-1$, $\delta\sigma$ is a dominant decoupled (en)
  source on superhorizon scales, which means that
\be
\delta\sigma=\delta\sigma^{(ad)}+\delta\sigma^{(en)}\equiv\delta\sigma^{(en)} \qquad \Rightarrow \delta\sigma^{(ad)}(\zeta_n,\,{\cal R}_{\pi_0})= 0 \,;
\ee 
\item When $c_b^2=0$, ${\cal R}$ is a dominant decoupled ($ad$) source
  on superhorizon scales,
  which means that
\be
{\cal R}={\cal R}^{(ad)}+{\cal R}^{(en)}\equiv {\cal R}^{(ad)}\qquad \Rightarrow {\cal R}^{(en)}(\zeta_n,\,{\cal R}_{\pi_0})= 0 \,.
\ee 
\end{itemize}
\end{enumerate}
Following the above steps, one arrives at eq. (\ref{structure})  and,
in addition, the degeneracy is resolved  by
\be
 \begin{split}
 \label{split}
 & c_b^2=-1\,,\;\;\; \zeta_n^{(ad)}+\frac{1}{3}\,  \frac{c_0^2}{c_b^2}\, \frac{1}{\bar \varphi'} \left(\frac{\bar\varphi'}{{\cal H}}\, {\cal R}_{\pi_0}^{(ad)}\right)'=0\,.\\
 & c_b^2=0\,,\;\;\; {\cal R}_{\pi_0}^{(en)} \to 0\,. 
 \end{split}
\ee
 A degeneracy in the amplitude persists for
  $c_b^2=-1$ . In that case, by using eq. (\ref{split}), we get
  \be 
{\cal R}_{\pi_0}^{(ad)}=\frac{3 \, c_b^2}{c_0^2 \, (2-3\, c_b^2)}\,\zeta_n\,,\qquad {\cal R}_{\pi_0}^{(en)}={\cal R}_{\pi_0}-{\cal R}_{\pi_0}^{(ad)}\,.
\ee
Furthermore, being ${\cal R}_{\pi_0}^{(ad)}\sim c_0^{-2}\, \zeta_n$, for
small $c_{s\,2}$ one has
\be
 {\cal R}_{\pi_0}^{(ad)} \ll {\cal R}_{\pi_0}^{(en)} \approx {\cal R}_{\pi_0}\,.
\ee
As will be shown in the next section, 
$\zeta_n$ provides the seed for  adiabatic perturbations at the
beginning of radiation domination; as a result CMB
data~\cite{Akrami:2018odb} imply that its spectral index $n_s^{(ad)}$
has to be red tilted. Thus, when $c_b^2=0$,  ${\cal R}_{\pi_0}$ will
be red-tilted too. However, when $c_b^2=-1$, ${\cal R}_{\pi_0}$ has
two components with different spectral indices, one is still
$n_{s}^{(ad)}$  and the second is $n_{s}^{(en)}$. For small deviation
from $c_b^2=-1$, one can set 
\be 
c_b^2=-1+\delta_b\,,
\ee
and the deviation from $n_{s}^{(en)} =1$ turns out to
be 
\be
 n_{s}^{(en)}\approx 1+6\,(\delta_b-\epsilon)+\eta\,;
 \ee
which can be blue-tilted. The consequences for the secondary production of gravitational waves is rather interesting and studied in section \ref{G_W}.

\section{Reheating}
\label{reheating}

Once the seed of primordial perturbations is produced, it is important
to study how the Universe reheats and gets to the radiation domination
era. In single clock inflation, the hypothesis of the Weinberg theorem
are satisfied~\cite{Weinberg:2003sw} and the inflationary predictions
are largely independent of reheating, however this is not the case when
more then one field is present, as for solid and supersolid inflation,  where neither ${\cal R}$ nor $\zeta$ are conserved on super horizon scales and moreover ${\cal R} \neq \zeta$.
As a consequence of the presence of $\varphi^0$, the pressure
perturbation is not proportional to $\delta \rho$
\be
\delta p = c_s^2 \, \delta p + \Gamma \, , \qquad \Gamma = \frac{\phi ' \left(c_b^2-c_s^2\right)}{a^4} \delta \sigma \, ;
\label{Gint}
\ee
thus the $\Gamma$ signals the presence of non-adiabatic
perturbations. 
Dealing with more than one component like in
$\Lambda$CDM, non-adiabaticity can also be present when the relative
energy density perturbations of two components are different:
$\delta_i\neq \delta_j$. The total non-adiabaticity
$\Gamma_{\text{tot}}$ contains both the intrinsic contribution for each
component of the form (\ref{Gint}) and the ``relative'' part
$\Gamma_{\text{rel}}$ that
takes into account that $\delta_i$ is not simply caused by the
``universal'' temperature perturbation. In the case of $\Lambda$CDM
with  a barotropic
equation of state for all the components only $\Gamma_{\text{rel}}$
is present and then $\Gamma_{\Lambda\text{CDM}}\equiv\Gamma_{\text{rel}}$; at superhorizon scales one gets
\be
\zeta= {\cal R}= \zeta_0 + \int_{a_{in}}^a
\frac{\Gamma_{\Lambda\text{CDM}}}{(1+w)\,\rho \, \tilde a} \,d \tilde a \, ;
\ee
where $\zeta_0$ is the adiabatic constant contribution. For a recent
discussion see~\cite{Celoria:2017xos}. 
\\
A pragmatic approach is to assume that reheating takes place 
instantaneously on a  time-like hypersurface  given in terms
of a 4-scalar $q$ as $q=$constant, or expanding at the linear order in
perturbation theory
\be
\bar q+\delta q= \text{constant} \, . 
\ee
A generic physical quantity ${\cal F}$ will be denoted by the subscript ${\cal F}_-$ when evaluated at the end of inflation, and with ${\cal F}_+$ when evaluated at the end of reheating. Thus, the change of ${\cal F}$
across $q$ will be simply written as  
\be
[{\cal F}]_\pm={\cal F}_+-{\cal F}_-\, ,
\ee
and the transition will be dictated by the Israel junction
conditions~\cite{Israel:1966rt}. By generalizing the results in
\cite{Deruelle:1995kd}, in the general gauge  (\ref{metric}), such
conditions read  at the linear level
\bea
&&\left[\zeta_{q}\right]_\pm  \equiv \left[-\Phi+{\cal H} \frac{\delta q}{\bar
    q'}\right]_\pm =0 \, ; \label{zetaq} \\
&& \left[\left(1-\frac{{\cal H}'}{{\cal H}^2}\right) \left({\cal R}- \zeta_{q}\right) \right]_\pm=0 \, ;
\label{Rzetad}\\
&&\left[ \Phi_{gi} \right]_\pm \equiv \left[\Phi+ {\cal
    H}\left(F-B'\right)\right]_\pm=0 \, .
\label{Phijump}
\ea
At the background level, the junction conditions imply that both $a$
and ${\cal H}$ are continuous on $q$.
The quantity $\zeta_q$ represents the gauge invariant curvature
perturbation of a constant $q$-hypersurface, and thus it is continuous
across $q$. From the transformation properties (\ref{transf}),
one can easily show that the junction conditions are gauge invariant. \\
As a reasonable assumption, we will take  $q$ to be the
particle number density $n$.
Intuitively, in the approximation of an instantaneous reheating, the rate for any channel  for the decay of inflatons  into a particle A becomes very large and the decay itself is democratic, in the sense if $n_A$ is the number density of the particle A and $n$ is the total number density, then
\be
\frac{\delta n_A}{\bar n_A} \to \frac{r_A\, \delta n}{r_A\, \bar n}\equiv
\frac{\delta n}{ \bar n}\, ;
\ee    
from the above relation and the particle number conservation $\bar{n}' + 3 \,{\cal H} \, \bar n=0$ we have that
\be
\left[\frac{\delta n}{\bar n'} \right]_\pm=0\, .
\ee
In the flat gauge, where $\Phi=0$, such a condition is precisely (\ref{zetaq}) with $q= n$
\be
\left[\zeta_n\right]_\pm=0\,,
\label{zetaj}
\ee 
When the field $\varphi^0$ is absent, namely $M_0=M_1=0$
(solid inflation limit),  one is back to the standard case where reheating takes place at a constant energy density $\rho$ hypersurface like
in~\cite{Endlich:2012pz,Celoria:2019oiu}. 
The continuity of $\zeta_n$ can also be shown following the same lines
of~\cite{Celoria:2019oiu} by a  generalization of the procedure given
in~\cite{Mukhanov:991646}. By  using the definition of $\zeta_n$  and $\delta \sigma$ we have that
\be
\zeta = \zeta_n-\frac{\delta \sigma \; \phi '}{18 \, \plm^2 \, a^2 \,  (w+1)
  \mathcal{H}^2} \equiv \zeta_n +\frac{\Gamma_{\text{eff}}}{3\, \rho\, (1+w)\,
  c_s^2}\, .
\label{Gzeta}
\ee
By integrating by parts the relation which
gives ${\cal R}'$ and by using the time-time component of the Einstein
equations, see \cite{Celoria:2019oiu}, one gets
\be
\label{ein_jump}
\left[{\cal R}\right]_\pm= \left[\frac{1}{3}\frac{k^2 \Phi}{{\cal H}^2 \epsilon}+ \frac{\Gamma_{\text{eff}}}{3 \, \rho\, (1+w)\, c_{s}^2} \right]_\pm\,;
\ee 
where the effective intrinsic entropic perturbation $\Gamma_{\text{eff}}$ before/after reheating is defined as follows:
\be
 \text{Inflation:} \;\;\; \Gamma_{\text{eff}}=-\frac{\bar\varphi'}{a^4}\, c_{s}^2\, \delta\sigma\,, \qquad \text{Radiation:}\;\;\; \Gamma_{\text{eff}}= \Gamma_{\Lambda\text{CDM}}\,.
\ee
Then (\ref{ein_jump}), is equivalent to 
\be
\left[\zeta-\frac{\Gamma_{\text{eff}}}{3\, \rho\, (1+w)\,
    c_s^2}\right]_\pm=\left[\zeta_n\right]_\pm=0\,,
\label{zeta_jump}
\ee 
demonstrating our intuition (\ref{zetaj}).\\
By using (\ref{epsdef}) and (\ref{zetaj}) the second junction condition (\ref{Rzetad}) reads
\be
\label{LCDM}
{\cal R}_+=\zeta_n+\frac{\epsilon}{\epsilon_+} \left({\cal R}_- -
  \zeta_n\right)\, , \qquad \qquad \qquad \epsilon_+= \left(1- \frac{{\cal
      H}'}{{\cal H}^2} \right)_{|\text{radiation}} \approx 2 \, ,
\ee
Let us consider the most important case where, after the Universe
reheats, a vanilla $\Lambda$CDM radiation dominated era is reached,
for which at superhorizon scales
\be
 {\cal R}_+= \zeta_+\,.
\ee
From the above relation and by using (\ref{LCDM}), the jump of $\zeta$
across the reheating hypersurface is
\be
\left[\zeta\right]_\pm={\cal
  R}_+-\zeta_-=\left(\zeta_n-\zeta_-\right)+\frac{\epsilon}{\epsilon_+}
\left({\cal R}_--\zeta_n\right) \, ;
\label{zetajs}
\ee
where, being $\zeta_n$ continuous, $\zeta_n{}_+=\zeta_n{}_-$ and 
has been denoted simply by $\zeta_n$. \\
Finally, one can calculate the total amount of non-adiabaticity
$\Gamma_{\Lambda\text{CDM}}$ present at the beginning of  the radiation
era. Indeed, by comparison with (\ref{zeta_jump})
\be
\left[\frac{\Gamma_{\text{eff}}} {3\, \rho\, (1+w)\,
  c_s^2} \right]_\pm = \frac{\Gamma_{\Lambda\text{CDM}}}{3\, \rho\, (1+w)\,
  c_s^2}_{|\text{Rad}}  -\;  \frac{\Gamma_{\text{eff}}} {3\, \rho\, (1+w)\,
  c_s^2}_{|\text{Infl}} = \left[\zeta \right]_\pm
\ee
The jump of $\zeta$ is given by (\ref{zetajs}), thus 
\be
\Gamma_{\Lambda\text{CDM}} =\frac{4}{3}  \, \rho_{\text{Rad}} \left[\frac{\Gamma_{\text{eff}}} {3\, \rho\, (1+w)\,
  c_s^2}_{|\text{Infl}}  + \left(\zeta_n-\zeta_-\right)+\frac{\epsilon}{\epsilon_+}
\left({\cal R}_--\zeta_n\right) \right]  \nb\\
= \frac{2}{3}  \, \rho_{\text{Rad}} \, 
\epsilon\, \left({\cal R}_--\zeta_n\right) \, ;
\ee
finally, taking into account that the above relation refers to superhorizon 
scales, from the results of Appendix \ref{gi-app} and \ref{can-app} we
arrive at 
\bea
\Gamma_{\Lambda\text{CDM}}&=&
 \frac{2}{3}  \, \rho_{\text{Rad}} \,  \epsilon \, \left[{\cal H}(1+c_1^2)
   \pi_L'-c_1^2 {\cal R}_{\pi_0}-\zeta_n\right] \nb \\
 &\approx &- \frac{2}{3}\,
 \rho_{\text{Rad}} \,  \epsilon \, \sum_{j=1}^2\left[\zeta_n^{(j)}\left(1+c_{s\,j}^2 (1+c_1^2)\right)+c_1^2 \, {\cal R}_{\pi_0}^{(j)} \right]\, .
\label{Gval}
\ea
Note that the contribution to the transmitted $\Gamma$ stays small  for small $c_{s2}$. Indeed, from eqs. (\ref{c_01_as}), we get
  that 
\begin{equation} 
  \epsilon \, c_1^2\, {\cal R}_{\pi_0}^{(2)} \propto  \begin{cases}
\epsilon\,
c_{s2}^2& c_b^2=-1 \\
 \epsilon\,
c_{s2}^{\frac{1}{2}}    & c_b^2=0
 \end{cases} \, .
\end{equation}
There is still a point to address. Take a generic field $X$ that satisfies a second order evolution equation with two independent solutions: one $X_{cg}$, growing or constant with scale factor $a$, and a second one $X_d$ decreasing with $a$. Clearly, the physically relevant solution is $X_{cg}$; however, even if the junction conditions prescribe that $\left[X\right]=0$, the constant/growing mode alone
can be discontinuous.
A classic example is given by the gauge invariant  Bardeen
potential $\Phi_{gi}$, which according to (\ref{Phijump}) is continuous
in the transition at constant $\rho$  with a sudden change of equation
of state in $\Lambda$CDM; however, from the continuity of $\zeta$ constant mode, one gets
\be
\label{Phi_jump}
 \Phi_{gi\,+} \mid_{\text{constant mode}}=\Phi_{gi\,-} \mid_{\text{constant mode}} \, \frac{(1+w_+)}{(1+w_-)}\frac{(5+3 \, w_-)}{(5+3\, w_+)}\,.
\ee  
Things are different in our non-adiabatic case. A clear understanding of the behavior of $\Phi_{gi}$  constant mode is crucial to predict the correct back reaction of tensor modes during radiation domination. Indeed, the validity of (\ref{Phi_jump}) crucially implies that the  $\Phi_{gi}$ gains a
factor $\epsilon^{-1}$ entering radiation domination.  
For simplicity, in the rest of this section we will work in Newtonian gauge, where $\Phi_{gi}$ coincides with $\Phi$. 
For each classic scalar field, it is convenient to distinguish among constant, decaying (absent during inflation) and entropic
(particular solution proportional to the non-adiabatic source term
proportional to $\Gamma$) modes. Once the decaying modes are under control, in principle, a reshuffling of constant and entropic
modes in the junction conditions is still possible.
Focusing on the entropic source  $\Gamma_{\Lambda \text{CDM}}$
relative to $\Lambda$CDM  where dark energy is just a cosmological
constant; neglecting baryons during  radiation domination, we have
two fluids: dark matter and photons as discussed
in~\cite{Celoria:2017xos} and $\Gamma_{\Lambda\text{CDM}}$ assumes the
form
\be
\Gamma_{\Lambda\text{CDM}}=\frac{8\, H_0^2 \, M_{pl}^2\, \Omega_m \,
  \Omega_r}{a^3\, \left(4 \, \Omega_r+3\, a\, \Omega_m\right)}\, s_0
(k)\,;
\label{Gvalm}
 \ee
with  $s_0(k)$ a scale dependent constant that is determined by using
(\ref{Gval}) at $t_+=-t_-$ \footnote{The equality $t_+=-t_-$ comes from the continuity of the Hubble conformal parameter $\left[{\cal H}\right]=0$.}.  
At superhorizon scales, the non-adiabatic contribution to $\zeta$ reads
\be
\zeta |_{en}=s_0 (k) \, \frac{a\,\Omega_m}{3 \, a\, \Omega_m+ 4\,
  \Omega_r}\, ,
\ee 
while the contribution to $\zeta_n$ is
\be
 \zeta_n |_{en}= \zeta|_{en} - \frac{\Gamma_{\Lambda\text{CDM}}}{3\,
   (\rho+p)\, c_s^2} \equiv 0\, .
\ee
Thus, during inflation $\zeta_n$ acts always as a source term for ${\cal \zeta}$, ${\cal R}$ and $\delta \sigma$ when $-1 \leq c_b^2 \leq 0$. The same exactly happens during the radiation domination where any entropic
contribution to $\zeta_n$ is compensated by an opposite contribution from $\zeta$ or ${\cal R}$ leading to 
\be
\left[\zeta_n \right] |_{\text{constant modes}}\equiv 0\,. 
\ee 
Following \cite{Mukhanov:991646}, expressing $\zeta_n$ in terms of the
Bardeen potentials in the Newtonian gauge, we get 
\be
\label{zeta_n_def}
\zeta_n=-\Phi-\frac{2 (\Phi'+{\cal H}\, \Psi)}{3\, {\cal H}\, (1+w)}-\frac{2\, k^2 \, \Phi}{9\,(1+w)\,{\cal H}^2} -\frac{\Gamma_{\text{eff}}}{3\, \rho\, (1+w)\, c_{s}^2}\,,
\ee
eq. (\ref{Phi_jump}) is non longer valid.
Imposing that 
\be
\zeta_n |_{en}=0\,, 
\ee
we get
\be
\Phi|_{en}+\frac{2 (\Phi'+{\cal H}\, \Phi)}{3\, {\cal H}\, (1+w)}|_{en}+\frac{\Gamma_{\Lambda\text{CDM}}}{3\, \rho\, (1+w)\, c_{s}^2} =0\,.
\ee
Considering the early stages of radiation domination, where dark
energy is negligible, we have
\be
 \Phi_{\text{rad}} \to -\frac{s_0(k)}{5 \,a^3}\, \left[a^3-2\, a^2\, a_e+8\, a\,a_e^2+16 \, a_e^3 \right]\,,
\ee
where $a_{eq}=\frac{\Omega_r}{\Omega_m}$ is the scale factor at the
matter radiation equality and we normalized the today's  scale factor
as  $a_0=1$. The same results could have been
obtained by directly solving the $\Phi$ equation of motion or
equivalently by expressing $\zeta$ in terms of $\Phi$ and $\Phi'$ and
enforcing that $\zeta_n$ is not affected by non-adiabatic perturbations.
Thus, eliminating $\Gamma_{\text{eff}}$ from (\ref{zeta_n_def}), we can extract the constant $\Phi$ mode during  the radiation phase 
\be
\label{correct}
\Phi |_{\text{constant mode}}=-\frac{3\, (1+w)}{5+3 \, w} \,
\zeta_{n}\,, \qquad w \to \frac{1}{3}\, ,
\ee
which is similar to the standard result with $\zeta$ replaced by
$\zeta_n$. The result (\ref{correct}) is not compatible with
eq. (\ref{Phi_jump}) that would imply the transmission of the ${\cal
  R}_{\pi_0}$ in the constant mode of $\Phi$. Thus, $\Phi$ gets an
enhancement of order $\epsilon^{-1}$ when the Universe transits into
the radiation era, without a further enhancement due to the presence
of ${\cal R}_{\pi_0}$ during inflation.
Summarizing, in the case of an instantaneous reheating, $\zeta_n$
determines initial conditions at superhorizon scales for the standard
evolution for the $\Lambda$CDM scenario with small deviations from a perfectly adiabatic spectrum of primordial perturbations.

\section{Primordial Non Gaussianity: a preview}
Primordial Non-Gaussianity (NG) is an essential tool to distinguish among different models of inflation. Single field inflation with its characteristic symmetry breaking pattern gives a small amount NG in the scalar and tensor sector, with the scalar part peaked in the local shape. A complete analysis of NG in supersolid inflation will be given in a companion paper~\cite{NGSS}, here we will outline some of the results needed to study the secondary production of GWs.
Given the presence of two scalars and tensor fields, the full cubic
action for a supersolid is quite complicated. Cubic terms can involve
three scalars (SSS), one scalar and two tensors (TTS), two scalars and
one tensor (TSS) and  three tensors (TTT); each contribution to the cubic
Lagrangian ${\cal L}^{(3)} $ in Fourier representation has the following general structure 
\be\label{L3}
{\cal L}^{(3)} \sim \omega\,M_{\text{pl}}^2 \, H^2 \, a^m\, D_{k\,k'\,k''} \, \xi_{1,\,k}\, \xi_{2,\,k'}\,\xi_{3,\,k''}\,, 
\ee     
where $\omega$ is a constant that sets the overall size of the vertex and $m$ determines its time evolution in terms of the scale
factor $a$; finally $D$ is a dimensionless function of the momenta and is determined by the structure of spatial derivatives acting on the fields entering the  vertex denoted by  $\xi_{i, k}$ which can be any combination of $\zeta_{n}$, $H^{-1}\,\zeta_{n}'$, ${\cal R}_{\pi_0}$, $H^{-1}\,{\cal R}_{\pi_0}'$, and
$h^s$; $h^s$ is the spin two tensor field (indices are omitted).
In general, one can show that 
\be
f_{\text{NL}} \sim \frac{\omega }{\epsilon }\,. 
\ee
The value of $\omega$ is determined by the relative size of the
derivatives of the Lagrangian density $U$ of the scalar sector with
respect to the rotational invariant independent operators.
In \cite{Endlich:2012pz} it was assumed the presence of a partial
cancellation among the derivatives of $U$ such that, even in
slow-roll, $\omega \sim 1$. Such extreme choice maximizes the
deviation from single field inflation, pumping up local NG to
$f_{\text{NL}} \sim \epsilon^{-1}$ which is in trouble with  recent
Plank constraints~\cite{Akrami:2019izv}. Here we take a more
conservative approach, considering that each derivative of $U$ is of order $\epsilon$ in
slow-roll expansion, leading to 
\be
\omega= \alpha \, \epsilon \, , 
\ee
with $\alpha$ an order one quantity. As a result, we get that
$f_{\text{NL}} \sim O(\epsilon^0)$ and, in addition, the cutoff of the
effective field theory describing a supersolid is higher.
Compared with NG in solid inflation, the presence of an additional
scalar introduces non-adiabatic perturbations controlled by the parameter $c_b^2$. This parameter has an important effect on any 3-point function involving ${\cal R}_{\pi_0}$ and, as we have seen, on the PS of ${\cal R}_{\pi_0}$ itself as discussed in section \ref{quant-sect}. \\
In particular, when $c_b^2 \to 0$, we can show that  the local $f_{\text{NL}}$ tends to be unacceptably {\it big } and strongly scale-dependent, unless some rather unnatural tuning is made.
As a result, when primordial NG is considered, the best choice is to
take $c_b^2 \approx -1$.
As it will be shown in the next section, in this case, supersolid inflation
features a rather exciting boost of the secondary gravitation waves production during inflation thanks to the cubic mixed TSS that is promising for future experiments.

\section{Gravitational Waves}
\label{G_W}

Given the current experimental upper bound on the tensor to scalar
ratio $r \le 0.5$, it is important to discriminate among different
inflationary models by telling how close to the limit the prediction for  $r$ can be. Indeed,  in the next few years, we will be able to probe the region $\in \left(10^{-1} \, \div\, 10^{-2}\right)$. 
Our analysis is similar to the one in \cite{Biagetti:2014asa}, where
secondary gravitational waves generated by a spectator scalar field
was studied. However, in that specific case, taking into account the related secondary scalar PS, considerably reduces the ratio $r$~\cite{Fujita:2014oba}.
On the contrary, in our supersolid model of inflation, the dominant cubic scalar vertex (SSS) is essentially unrelated to the dominant
tensor-scalar-scalar (TSS) cubic one. That gives us room to effectively
enhance $r$ to get close to its experimental upper limit with only the secondary tensor production.
That feature singles out supersolid from single field inflationary
models where the dominant GW production is not very sensitive to NG
and  gravitational waves back-reaction is much smaller than the one
generated during the radiation phase as it was observed originally in~\cite{Matarrese:1992rp, Matarrese_1994,Mollerach_2004} and later extended in~\cite{Ananda:2006af, Baumann:2007zm}. 
  
Spin two tensor perturbations are defined by
\be
g_{ij} = a^2\,
\left(\delta_{ij} + h_{ij}\right) \, , \qquad \qquad \delta_{ij} h_{ij}
=\de_j h_{ij} = 0 \, ;
\ee
where $h_{ij}$ is the transverse and traceless part of the metric tensor. During Inflation, the corresponding  quadratic/cubic Lagrangian can be written as 
\be 
 L_{T}= M_{pl}^2\,a^2  \left[  \frac{1}{2}\left(h_{ij}{}'
     \,h_{ij}{}'-(M_2\, a^2-\partial^2)\,  h_{ij} h_{ij}\right)  +
   h_{ij} \, {\cal S}_{ij} \right]\,,
\ee  
where $S_{ij}$ is a transverse-traceless  quadratic-source term. The evolution equation for GWs is
\be
\label{eqh}
h_{ij}''+2\;{\cal H}\;h_{ij}'-\Delta\;h_{ij}={\cal S}_{ij} \, .
\ee
where we neglect the mass $M_2$ being proportional to $\epsilon$, see
(\ref{par}).
The leading contribution to ${\cal S}_{ij}$ comes from the cubic interaction terms containing one spin two field $h_{ij}$
and two scalars. There is 
a  ``universal'' contribution from cubic terms  in the
Einstein-Hilbert Lagrangian and a graviton scalar interactions in the ``matter'' sector; namely   ${\cal S}_{ij}={\cal S}^{(EH)}_{ij}+{\cal S}^{(\text{Matter})}_{ij}$.
The leading structure of the EH interactions comes from derivatives
of scalar perturbations and has the following structure 
\be
\begin{split}
\, {\cal S}^{(EH)}_{ij} \propto \;\partial_i \Phi_{gi}  \,\partial_j  \Phi_{gi}
\, .
\end{split}
\ee
The matter contribution ${\cal S}^{(\text{Matter})}_{ij}$ changes effectively during the universe evolution
\footnote{During matter/radiation domination (Matter=Matter/Radiation Fluid), with DM/photons represented as a perfect fluid, the source term becomes
 \be
 {\cal S}^{(\text{Rad})}_{ } \propto{ \frac{1}{\bar \rho+\bar
   p}}\left(\partial\Phi_{gi}'\,\partial\Phi_{gi}'+{\cal
     H}\;\partial\Phi_{gi}\,\partial\Phi_{gi}'+{\cal
     H}^2\;\partial\Phi_{gi}\,\partial\Phi_{gi} \right) \, .
  \ee.}.
In our case, during the inflationary period (where Matter = Inflaton), 
among all the possible TSS vertices, the  dominant one is given by the following cubic lagrangian (see the structure in (\ref{L3}))
\be
{\cal L}^{(3)}_{TSS}=-\epsilon\,\alpha\,a^2\;M_{pl}^2\;k'^{i}\;k''^{j}\;h_{ij}(k)\;{\cal R}_{\pi_0}(k')\;{\cal R}_{\pi_0}(k'')\qquad \to\quad
{\cal S}^{(\text{Infl})}_{ij} = \epsilon\,\alpha\,a^2\, \partial_i {\cal
  R}_{\pi_0}\, \partial_j  {\cal R}_{\pi_0} \, ,
\label{TSSc}
 \ee 
with $\alpha$ a constant given by
\begin{equation}
\alpha= 2\,\left(c_{2\, w}^2-c_1^2\right)\, ;
\end{equation}
where
\be
\begin{split}
 c_2^2 \equiv c_{2\,\tau}^2 +&c_{2\,w}^2;\qquad
   c_{2\,\tau}^2  =-\frac{a^2}{9 \, {\cal H}^2
   \, \epsilon}\left(U_{\tau_y}+U_{\tau_z} \right)\,; \qquad
 c_{2\,w}^2= -\frac{ a^2}{9 \, {\cal H}^2
   \, \epsilon}\left(U_{w_y}+U_{w_z}\right)\, ;
\end{split}
\ee 
with $c_{2\, w}^2$ the part of $M_2$~\footnote{See Appendix \ref{mass-app}.} proportional to the derivatives of $U$
with respect to the operators $\{ w_i\}$.  \\
The analysis of the role of the  operators $w_i$ is interesting. At the zero and  first order in the perturbation theory
 they are degenerate with the  operators $\{\tau_i\}$, so are sensitive   only to the solid structure of the medium.
It is only at second order that the $\{w_i\}$ start to discriminate a solid from a  supersolid.
From the structure of the  $\{w_i\}$, we see that the $\varphi^0$ scalar field, related to the superfluid part, is intrinsically coupled to the $\varphi^a$ fields, describing the solid side.
Thus, while the presence of the operators $\{ w_i\}$ is immaterial at the
linear level, it plays an important role for non-Gaussianity.
If the operators $\{w_i \}$ are absent,  automatically $c_{2\,w}=0$
and $\alpha=-2\, c_1^2$. However,  from (\ref{c_01_as}), in this case
\begin{equation*}
\alpha \sim c_{s2}^{5} \ll 1
\end{equation*}
and TSS vertex is negligible.
The  only constraint on $c_{2\, w}^2$ comes from stability: $c_L^2 >
2\, c_0^2 \, c_b^4$ and $c_{s\,2}^2<c_L^2<c_{s1}^2$.\\
Given the presence of  ${\cal
  R}_{\pi_0}$, the size of the source is very sensitive to the value of  $c_{s2}$
(typically $\propto c_{s2}^{-3}$).
In our specific case, during inflation, we get that the Einstein Hilbert term is always suppressed 
${\cal S}^{(\text{Infl})}\gg{\cal S}^{(EH)} $, while during the
radiation phase nothing more than what is described
in~\cite{Ananda:2006af, Baumann:2007zm} happens; the only difference
is that the Bardeen potential is proportional to $\zeta_n$ instead of
$\zeta$,  see (\ref{correct}). 
The tensor PS has two contributions: one (primary PS)  ${\cal P}_{h}^{(1)}$ from the quantum fluctuations during the dS period and calculated with the homogeneous quadratic action of $h_{ij}$, and another classical contribution (secondary PS) ${\cal P}_{h}^{(2)}$ coming from the interactions of $h_{ij}$ with the other scalar fluctuations.
This last term can be calculated by finding the particular solution of (\ref{eqh}) proportional to ${\cal S}^{(Matter)}$.
The computation of the primary tensor PS is  standard;
denoting with $H_i$ the Hubble parameter during the dS phase,  in the case $c_b^2=-1$  we have 
\be
{\cal P}_{h}^{(1)}=
\frac{H_i^2}{8\, M_{pl}^2 \,\pi^2}\, (-t\,k)^{\frac{8}{3}\, c_2^2 \, \epsilon}\, k^{2 \,c_L^2 \,\epsilon }\, 
\simeq \epsilon\;\frac{{\cal P}_{\zeta_n}}{\bar{\cal P}}. 
\ee
Remember that $\zeta_n$ fluctuations represent the primordial seed for scalar perturbations during the radiation phase. The particular solution { of (\ref{eqh})} can be obtained by using the Green function ($g_k (t,\,t')$) method
\be
\begin{split}
& h(t,\,k)=
\int_{-\infty}^t g_k(t,\,t')\,{\cal S}_{k}(t')\, dt'\,, \qquad {\cal
  S}= \varepsilon_{ij}^{(s)\,*}\, {\cal S}_{ij}\,, \\[.2cm]
&\;\; \partial_t^2\,g_k (t,\,t') + 2 \, {\cal H}\;\partial_t\,g_k (t,\,t')+k^2\,g_k (t,\,t')=\delta(t-t')\,. 
\end{split}
\label{partsol}
\ee
{  The above solution can be used to extract the PS ${\cal P}_{h}^{(2)}$ for the secondary production of GWs as} 
\be
  (2\,\pi)^3\, P_{h}^{(2)}(k\,t)\, \delta(\textbf{k}-\textbf{p}) \equiv   \int_{-\infty}^t \int_{-\infty}^t\, dt'\,dt'' \, g_k(t,\,t')\, g_p(t,\,t'')\, \langle S_k (t')\,S_p (t'') \rangle \,,
\ee
where $\langle S_k (t')\,S_p (t') \rangle$ represents a Gaussian
4-point scalar correlator.
During the inflationary period, the above correlator is
proportional to $\langle {\cal R}_{\pi_0}^4\rangle$; in the limit
of a small $c_{s2}$  one gets the following  estimate for the secondary scale-invariant PS
\be
\label{sec_prod} 
 {\cal P}_h^{(2)} \approx \alpha^2\, \epsilon^2\,\frac{2\,
   \pi^2}{c_{s2}}\,{\cal P}_{{\cal R}_{\pi_0} }^2\,
 = 
  \alpha^2\, \epsilon^2\,\gamma^2\;\frac{2\,
    \pi^2}{c_{s2}^{13}\;\bar{\cal P}^2}\,{\cal P}_{\zeta_n}^2 \, \, ,
\ee
where we have defined $\gamma$ such that
\be
{\cal P}_{{\cal R}_{\pi_0} }=\gamma\;\frac{{\cal P}_{\zeta_n
  }}{c_{s2}^6\;\bar{\cal P}} \, .
\ee
The final expression for the total tensor  PS is given by
\be
{\cal P}_h ={\cal P}_h^{(1)}+{\cal P}_h^{(2)}={\cal P}_{\zeta_n}\;\frac{\epsilon}{\bar{\cal P}}
\left(1+ \alpha^2\, \epsilon\,\gamma^2\;\frac{2\,
    \pi^2}{c_{s2}^{13}\;\bar{\cal P}}\,{\cal P}_{\zeta_n} \right)
\, .
\ee
The presence of the coupling constant (\ref{TSSc}) which controls the TSS vertex gives rise to the question of whether the cubic scalar
interactions can give sizeable contributions to the scalar PS.
The SSS dominant interaction Lagrangian for the scalar one loop corrections to ${\cal P}_{\zeta_n}$  has  the following  structure (\ref{L3})
\be
{\cal L}^{(3)}_{SSS}=-\epsilon \,a^2\,M_{\text{pl}}^2\;\left[\beta_2 \; k'^{i}\,k''^{i}\;+\beta_1 \;\frac{k^i\,k^j\;k'^{i}\,k''^{j}}{k^2}
\right] \,\,    \zeta_n(k)\;{\cal R}_{\pi_0}(k')\;{\cal R}_{\pi_0}(k'')
 ,
\ee
and when $c_b^2 \approx -1$
\be
\beta_1= 12\, c_{2\,w}^2\,, \qquad \beta_2=-\frac{1}{2}\left\lbrace 3\, \frac{\sigma}{M_{\text{pl}}^2\,H^2\,\epsilon}-4\, \left[ c_{2\,w}^2+3\,\left(c_0^2\,c_b^2+\frac{U_{by}}{8\, H^2\, \epsilon}\right)\right]\right\rbrace\,.
\ee
As usual, $U_{by}$ and $\sigma$ will be taken to be of order $\epsilon$ to get $\beta_i$ order one, as discussed in the previous section. Furthermore, while $\beta_1$ vanishes in the absence of
$w$-operators, $\beta_2$ is generically different from zero.
For generic $\beta_i \sim 1$, the computation of the non-linear
correction to the scalar PS is complicated. A reasonable estimate is given by
\be
\begin{split}
 &{\cal P}_{\zeta_n }^{(2)}\propto\epsilon^2\,\beta_i^2\, \frac{2\,\pi^2}{c_{s_2}}\,
 {\cal P}_{{\cal R}_{\pi_0}}^2=\epsilon^2\,\beta_i^2\,\gamma^2 \frac{2\,\pi^2}{c_{s_2}^{13}\;\bar{\cal P}^2}\,
 {\cal P}_{ \zeta_n}^2 \, ,
 \end{split}
\ee
with the total  scalar PS given by ${\cal P}_{\zeta_n }+{\cal P}_{\zeta_n }^{(2)}$.
The possibility to have a regime where the secondary tensor production
is dominant while the secondary scalar contribution is negligible, namely 
\be
\frac{{\cal P}_{h }^{(2)}}{{\cal P}_{h}^{(1)}}\gg1 \, ,  \qquad
\frac{{\cal P}_{\zeta_n }^{(2)}}{{\cal P}_{\zeta_n}}\ll1 \, , 
\ee
gives
\be
\epsilon\ll \frac{\alpha^2}{\beta^2}\,\bar{\cal P},\qquad 
\frac{2\,\pi^2\,\gamma^2\,{\cal P}_{\zeta_n}}{\bar{\cal P}}\,\frac{\beta^2\,\epsilon^2}{\bar{\cal P}}\ll c_{s2}^{13}\ll
\frac{2\,\pi^2\,\gamma^2\,{\cal P}_{\zeta_n}}{\bar{\cal
    P}}\,\alpha^2\,\epsilon \, .
\ee
Taking $\gamma\sim1$, ${\cal P}_{\zeta_n}=10^{-9}$ and $\bar{\cal P}=32$ we get
\be
\label{con1}
0.15\;\left(\beta^2\,\epsilon^{2}\right)^{1/13}\ll c_{s2}\ll
0.2\;\left(\alpha^2\,\epsilon\right)^{1/13} \, .
\ee
Relation (\ref{con1}) is valid whether or not
  the operator $w_i$ are present. However, when $c_{2\, w}^2$ is zero,  the inequality 
\be 
\beta^2 \epsilon^2 <  \alpha^2 \,\epsilon \sim {c_{s2}}^{10}\,\epsilon\,,
\ee
is valid only if $\epsilon$ is very small, and then is much more
tuned.  The presence of the parameter $c_{2\, w}^2$ makes the
gravitational waves secondary production dominant for a suitable region of the parameters space, even if the ${\cal R}_{\pi_0}$ is not efficiently transmitted in the scalar sector after inflation.
Even the case where the secondary scalar and tensor production
are both relevant is interesting.
A rough estimate gives 
\be
r=\frac{{\cal P}_{h }^{(2)}}{{\cal P}_{\zeta_n
  }^{(2)}}=\frac{\beta_i^2}{\alpha^2} \, ,
\ee
which is very sensitive to the detail of the non-linear structure of the theory.\\
Let us mention  that, even if a cubic coupling between tensors and transverse vectors of the schematic form $h_{ij} (\pi_i' \pi_j'+\de \pi_i \de \pi_j)$ exists, we do not expect an enhancement similar to the one found due to the scalars. The transverse vector sector is very similar to solid inflation, and in the limit of small $c_{s\,2}$
\begin{equation*}
\langle \pi_i\, \pi_i \rangle \sim O(c_{s\,2}{}^0)\,,  
\end{equation*}
with $\pi_i$   defined by the linear theory. 
Thus, the secondary production of GWs from the vector
  sector  is much smaller than the one from the scalar sector which is of  order  ${\cal P}_{{\cal R}_{\pi_0}}{}^2$. \\
All the above expressions are given at the leading order in a slow-roll expansion and the PS are scale-free, modulo small slow-roll corrections.\\
Finally, let us  estimate the tilt of secondary GWs production.
The next to leading   corrections to the primordial
tensor PS  can  be obtained by  simply substituting ${\cal P}_{{\cal R}_{\pi_0}}$ in eq. (\ref{sec_prod}) with the complete expression $k^3\,|{\cal  R}_{\pi_0}|^2 /(2\, \pi^2)$. In the $c_b^2=-1$ case, we have three contributions with three different tilts (see eq. (\ref{structure})): $n_s^{(ad)}$, $n_s^{(en)}$, and  
\begin{equation}
n_s^{(ad-en)}=\frac{n_s^{(ad)}+n_s^{(en)}}{2}\,.
\end{equation} 
As we argued, there is the possibility to get a
blue-tilted index for $n_s^{(en)}$ tilt, and being $n_s^{(ad)}$
red-tilted, the ($en$) index term will be the dominant one for scales much
smaller than the CMB ones~\footnote{The standard CMB-like  pivot scale is 
  $k^*=0.002\, \text{Mpc}^{-1}$.},
\be
 {\cal P}_{{\cal R}_{\pi_0}} \to  {\cal P}_{{\cal R}_{\pi_0}}^{(en)}\,(-H\,t)^{6\,\delta_b-5\,\epsilon+\eta}\, k^{n_s^{(en)}-1} \,.
\ee 
Thus, eq. (\ref{sec_prod}) reduces to
\be
{\cal P}_h^{(2)} \approx \alpha^2 \, \epsilon_i^2 \, \frac{2\, \pi^2}{c_{s2}}\, {\cal P}_{{\cal R}_{\pi_0}}^{(en)}{}^2 \,(-H\,t)^{12\,\delta_b-10\,\epsilon}\; k^{2(n_s^{(en)}-1)}\,,
\ee
then
\be
n_T= 2\,n_s^{(en)}-1\equiv 1+12 \, (\delta_b-\epsilon_i)+2\, \eta\,,
\ee
which is blue tilte when 
\be
\delta_b > \epsilon_i-\frac{\eta}{6}\,. 
\ee
The presence of a blue $n_T$ parameter will be an interesting tool to
test inflationary models in future high sensitivity experiments of
Gravitational waves detection\cite{amaroseoane2017laser}. Indeed, for
modes  that re-enter  the horizon during radiation domination, the GWs energy density spectrum\cite{Boyle_2008, Smith_2019}  goes as
\be 
\Omega_{GW} \propto k^{n_T-1}\,,
\ee
if $n_T-1>0$, it could grow up to the frequencies $f=c\, k/2 \pi $ in
the milli-Hertz band, where we expect the maximum of LISA sensitivity\cite{Bartolo:2016ami,Guzzetti:2016mkm}.
\\

\section{Conclusion}
\label{con}
An extreme synthesis of the production of the seeds for cosmological perturbations in supersolid inflation is given in
  Table 4 where the magnitude and the fate of the leading order power spectra of scalar and tensorial perturbations are shown. 
\begin{table}[H]
\begin{center}
\begin{tabular}{|c|}
\hline
\\
$\quad\underbrace{{\cal P}^{(1)}_{en}\sim\left( \frac{{\cal P}^{(1)}_{ad}}{c_{s2}^6}\right)
\quad\gg\quad
  {\cal P}^{(1)}_{ad/en}\sim\left( \frac{{\cal P}^{(1)}_{ad}}{c_{s2}^3}\right)}_{\rm Dissipated\,during\, the\, reheating}
  \quad\gg\quad
\underbrace{ {\cal P}^{(1)}_{ad}\sim10^{-9}  \quad\gg\quad{\cal P}^{(1)}_{h}\sim\epsilon \;{\cal P}^{(1)}_{ad} 
}_{\rm transmitted\,during\, the\, reheating}
\qquad$\\
\\
\hline
\end{tabular}
\caption{\small Relations among  power spectra (leading order)
  and their fate in a  instantaneous reheating, in the limit $c_{s2}\ll 1$.\\ 
    $ad=$  adiabatic   (${\cal P}^{(1)}_{ad} = {\cal P}^{(1)}_{\zeta_n} $) and $en=$  entropic (${\cal P}^{(1)}_{en} ={\cal P}^{(1)}_{{\cal R}_{\pi_0} }$)}.
\end{center}
\label{default}
\end{table}%
\no
Adiabatic  perturbations are related to the solid part of the medium ($\zeta_n\propto\pi_L\propto\varphi^i$).
 The presence of large entropic perturbations,  related to its
 superfluid component (${\cal R}_{\pi_0}\propto\pi_0\propto
 \varphi^0$), potentially can enhance the   PS of the other fields by next to leading corrections.\\
 The secondary production is generically suppressed for inflaton-like fields 
 \footnote{Inflaton perturbations in a quasi-deSitter background has its modes proportional to
 $H^2_i/(\epsilon\;M_{pl}^2)$. In the supersolid case we have two scalars ($\zeta_n $ and ${\cal R}_{\pi_0}$) and one transverse vector field 
$ \pi^i_T$. Transverse vectors decay subhorizon, we also expect  a
suppressed contribution to the secondary PS and will not be discussed
in this paper.},
tensor  perturbations  play the role of   spectator fields  (with a PS proportional to $H^2_i/( M_{pl}^2)$) and the interaction with entropic scalar perturbations enhances  considerably their secondary PS.
Indeed,  the following scenario is possible \footnote{Actually, in section 6 we get an extra $1/c_{s2}$ enhancing factor from phase space integration. 
\bea
 &&{\cal P}^{(2)}_{ad}\sim\epsilon^2\;\left( {\cal P}^{(1)}_{en} \right)^2\quad{\rm where }\qquad  
\frac{{\cal P}^{(2)}_{ad}}{{\cal P}^{(1)}_{ad}}\sim
 \;\left(\frac{\epsilon^2}{c_{s2}^{12}}\right)\;{\cal P}^{(1)}_{ad} \ll 1\,;\\
 &&
 {\cal P}^{(2)}_{h}\sim\epsilon^2\;\left(  {\cal P}^{(1)}_{en}\right)^2\quad{\rm where }\qquad \frac{{\cal P}^{(2)}_{h}}{{\cal P}^{(1)}_{h}}\sim
 \;\left(\frac{\epsilon }{c_{s2}^{12}}\right)\;{\cal P}^{(1)}_{ad}  \gg 1\,;
\ea 
${\cal P}^{(2)}_{ad}$ and $ {\cal P}^{(2)}_{h}$ refer to the next
 to leading contribution to the adiabatic scalar and tensor power spectra.
 The above consideration are valid in the case of a small $c_{s2}$. 
From the above results we can extract some general conclusions related to the
 pattern of symmetry breaking during inflation.} We have systematically explored the physical consequences of
the breaking of the full set of diffeomorphism of general relativity down
to $ISO(3)$. The breaking pattern is triggered by the background
configuration of four scalar fields and, in order to allow dS spacetime
as a solution, we have considered an additional set of internal symmetries comprising $SO(3)$ internal
rotations and four shift symmetries. The  four scalars $\varphi^A$ can be interpreted
as the coordinates of a supersolid embedded in spacetime and the
corresponding effective Lagrangian we have studied is the most general one consistent  with
the given symmetries at the leading order in a derivative
expansion. As a comparison, in the effective description of single
clock inflation~\cite{Cheung:2007st} the residual symmetry comprises three dimensional
diffeomorphism with one scalar and two tensor propagating modes, while
in our supersolid inflation, we have two scalars, two transverse vectors
and two tensors. Interestingly, as a benefit of the supersolid
interpretation, the scalar field fluctuations can be interpreted as
phonons modes and non-adiabatic perturbations.
Given the symmetry breaking pattern and the number of propagating
modes, the difference with single clock inflation are significant both at
the linear and non-linear levels.
At the linear level, the symmetry breaking pattern gives rise to a
peculiar kinetic mixing between the two scalars that makes the
quantization and the computation of the linear power spectra
non-trivial. A similar (but different) mixing is found in chromo-natural inflationary
models~\cite{Tolley_2010, Dimastrogiovanni_2013,
  Dimastrogiovanni:2012ew}, non-thermal production of gravitinos \cite{Nilles:2001fg}, multi-field
inflation \cite{Tolley_2010} and in effective theories of
inflation\cite{Bartolo:2015qvr}.  Our analysis and results differ
from the previous ones: we do not use perturbations theory  to resolve
the kinetic mixing but rely on Hamiltonian analysis and a set of
canonical transformations to reduced the dynamical system to two
uncoupled harmonic oscillators in the limit of large momentum $k$. As
a consequence, cross-correlations in the scalar power spectra are
unavoidable. The presence of the scalar $\varphi^0$ associated with the
superfluid component introduces the important parameter $c_b^2$ for the superhorizon evolution of the scalars. 
The hypothesis of the Weinberg theorem are explicitly violated. Indeed, we get both the presence of the anisotropic stress which is not negligible in the $k \to 0$ limit, and perturbations are non-adiabatic in general. Thus, neither the comoving curvature perturbation ${\cal R}$ nor the curvature perturbation $\zeta$ are conserved, moreover they differ on superhorizon scales, though their superhorizon evolution is only due to small slow-roll corrections. In the range $c_b^2 \in [-1, \, 0]$, all the relevant scalar power spectra are scale-free, modulo small slow-roll corrections, in agreement with experimental constraints. Because of the presence of
two scalar propagating
degrees of freedom, there is no smooth limit that leads to solid
inflation~\cite{Endlich:2012pz} and thus the predictions at the
level of linear power spectra are rather different. The system of coupled second
order differential equations for the linear evolution of the two
independent scalar perturbations are complicated enough due to the non-trivial kinetic mixing
to elude an analytical solution for a generic time $t$ unless
$c_b^2=0, \, -1$. { Luckily enough, these boundary values for
$c_b^2$ are such that the relevant power spectra are almost scale-free}.
Among the various scalar perturbations, we select the  power spectra
of the 
curvature perturbation $\zeta_n$ of the constant particle number $n$  hypersurface and
and curvature perturbation ${\cal R}_{\pi_0}$  of the
constant $\varphi^0$ hypersurface and the relative cross correlations,
studying in detail their properties as a function of $c_b^2$ and the
speed of sounds of the two independent diagonal scalar modes. 
In the instantaneous reheating approximation, by extending the analysis
in~\cite{Deruelle:1995kd}, we analyze how the seed of
primordial perturbations are transmitted to the standard hot radiation
dominated era of $\Lambda$CDM. Besides the standard adiabatic component, a small isocurvature part can be written as a linear combination of $\zeta_n$ and ${\cal R}_{\pi_0}$ evaluated at the end of inflation.
Also the prediction for primordial non-Gaussianity is rather
interesting; we leave a detailed account for a companion paper,
focusing on the secondary  production of gravitational waves during
inflation. The structure of the tensor-scalar-scalar cubic vertex is
such that it is possible to enhance the secondary production, saturating the experimental bound, still keeping the scalar bispectrum within the
limits set by Planck. Finally, the spectral index of
  GWs PS can be blue-tilted, enhancing the chance of a direct
  detection of the primordial stochastic background.\\
In conclusion, supersolid inflation is an interesting alternative to single clock inflation to explore different symmetry breaking patterns with a clear experimental signature.

\section*{Acknowledgements}
The work of DC and LP was supported in part by Grant No. 2017X7X85K 
``The dark universe: A synergic multimessenger approach" 
under the program PRIN 2017 funded by Ministero dell'Istruzione, 
Universit\`a e della Ricerca (MIUR). 

\begin{appendix}

 \section{Parameters ${\pmb{M_\alpha}}$}
  \label{mass-app}
The parameters $\left \{ M_\alpha \, ; \alpha=0,1,3,4 \right
\}$ entering in the quadratic action (\ref{quadflat})
are defined by the following derivatives of the Lagrangian density
around the background
\be
\bar g_{\mu \nu} = a(t)^2 \, \eta_{\mu \nu}\, , \qquad \bar \varphi^0
= \bar\varphi(t) \, , \qquad \bar \varphi^i
= x^i \, \qquad i=1,2,3 \, .
\ee
\bea
&& M_0 = \frac{\bar\varphi'{}^2 \left(U_{\chi \chi }+2 \, 
    U_{y \chi}+U_{yy}\right)}{2 a^2} \, , \qquad  M_1
=-\frac{\bar\varphi'{} \, U_{\chi }}{ a} \, , \qquad M_2 = -\frac{4}{9} \left(U_{w_Y}+U_{w_Z}+U_{\tau _Y}+U_{\tau
    _Z}\right) \, , \nb\\
&&  M_3 = \frac{M_2}{3}+\frac{1}{2} \,a^{-6}\,U_{bb} \, , \qquad M_4 = \frac{\bar\varphi ' \left[U_{b
       \chi}+U_{by} -a^3 \left(U_{\chi }+U_y\right)\right]}{2 a^4}\, ;
\ea
where all the derivatives are evaluated on the background values of
the operators by which $U$ depends on. The Minkowski background corresponds to $\bar\varphi'=a=1$.

\section{Gauge Invariant Operators and Perturbations}
\label{gi-app}
Being the background $SO(3)$ invariant, cosmological perturbations
can be decomposed in a scalar, vector, and tensor sector. In a generic gauge, scalar perturbations can be written as 
\bea
\label{metric}
&& \varphi^0 = \bar\varphi' + \pi_0 \, , \qquad
\varphi^i=x^i+\partial^i \pi_L+\pi_T^i \,, \\
&&g_{00}=-a^2 \, (1-2 \, \Psi)\, , \qquad g_{0i}=a^2 \, \partial_i F\,
, \qquad g_{ij}= a^2 \, \left[(1+2 \, \Phi)\,\delta_{ij}+2 \, \partial_{ij} B\right]\, .
\ena
Consider an infinitesimal coordinates transformation; in the scalar
sector we have that
\be
x^\mu \to \tilde{x}^\mu= x^\mu+\epsilon^\mu \, , \qquad \epsilon^\mu =
(\epsilon^0 , \, \partial^i \beta)\, . 
\ee
The scalar parts of metric and the perturbation of $\varphi^A$  transform according with~\footnote{The 
  variation $\Delta _{\text{gauge}} A$ of a quantity $A$ is defined as $\tilde
  A(x) -A(x)$ evaluated at the linear order in
  perturbation theory.}
\be
\begin{split}
&\Delta_{\text{gauge}} \Psi =\epsilon^0{}'+{\cal H}\, \epsilon^0\,,
\qquad \Delta _{\text{gauge}} F= \epsilon^0-\beta'\, , \qquad \Delta
_{\text{gauge}} \Phi=-{\cal H} \, \epsilon^0\, , \\
&\Delta _{\text{gauge}} B=-\beta \, , \; \; \qquad \qquad
\Delta _{\text{gauge}} \pi_0 = -\bar \varphi'\, \epsilon^0\,,\qquad  \Delta _{\text{gauge}} \pi_L =-\beta \, .
\end{split}
\label{transf}
\ee
From the above transformation properties, one can construct the
following gauge invariant perturbations
\be
\pi_{L,\,gi}=\pi_L-B\, , \qquad \qquad
\pi_{0,\,gi}= \pi_0 -\frac{\bar\varphi'}{{\cal H}} \Phi \, ;
\ee
and the corresponding curvature perturbations
\be
\zeta_n=\frac{k^2}{3} \pi_{L, \, gi}=-\Phi+\frac{{\cal
     H}}{\bar{n}'}\,\delta n \, , \qquad \qquad  
 {\cal R}_{\pi_0}=\frac{{\cal H}}{\bar\varphi'}\pi_{0,\,gi}= -\Phi+\frac{{\cal H}}{\bar\varphi'}\,\pi_0 \, .
 \label{Zetan_Rpi0}
  \ee
Together with
 \be
 \begin{split}
  &\zeta = \Phi + {\cal H} \frac{\delta \rho}{\bar{\rho}'} \, ;\\
  & {\cal R} = \Phi + {\cal H} \,v ,  
\end{split}
\ee
$\zeta_n$ and $ {\cal R}_{\pi_0}$ represent the fundamental gauge
invariant scalars.
While $\zeta_n$ represents the curvature of constant number density
hypersurfaces, ${\cal R}_{\pi_0}$ can be identified as the curvature
perturbation orthogonal to the velocity of the superfluid component in
the supersolid, see (\ref{svel}); whose spatial part, at the linear level is given  by 
\be
\nu^i= -\frac{1}{\bar\varphi'}\, \partial_i\pi_0\,. 
\ee 
From (\ref{sigmadef}), we have
\be
\delta\sigma = \frac{2 \, a^4  \, \plm^2 \, M_0}{\bar\varphi'} \left[\Psi +3 \, c_b^2\, \Phi+\frac{\pi_0'}{\bar\varphi'}+c_b^2\,k^2 \pi_{L,\, gi}\right]\, ,
\ee
and it is gauge invariant; taking the time derivative we arrive at
\be
\delta\sigma' = \frac{k^2 \, a^4\,  \plm^2 \, M_1}{\bar\varphi'} \left[\frac{\pi_0}{\bar\varphi'}-\pi_{L,\,gi}{}'+(F-B')\right]\, .
\label{sigmad}
\ee
Adiabatic media, solids for instance, are  characterized by  $u^\mu \de_\mu \sigma =0$ at the non-perturbative level. This is the case when the Lagrangian $U$ does not depend on $\chi$ in \cite{Celoria:2019oiu}. In particular, this implies that at the linearized level $M_1=0$, in perfect agreement with (\ref{sigmad}) which for such a class of media gives $\delta\sigma'=0$. 
The linearized EMT can be written as the perturbed EMT for a perfect
fluid plus an anisotropic stress contribution 
\be
T_\nu^\mu=(\bar{p}+\bar{\rho}+\delta p+\delta\rho)\;\bar{\mathfrak{U}}^\mu\;\bar{\mathfrak{U}}_\nu+(\bar{p}+\bar{\rho})\;(\delta
\mathfrak{U}^\mu\;\bar{\mathfrak{U}}_\nu+\bar{\mathfrak{U}}^\mu\;\delta
\mathfrak{U}_\nu)+ (\bar{p}+\delta p)\;\delta^\mu_\nu+\Pi^\mu_\nu \, ,
\label{emt}
\ee
where $\bar{\mathfrak{U}}_\mu=(-a,\vec{0})$ is the background 4-velocity, $\delta \rho,\;\delta p$ are the perturbations of energy density and
pressure. The anisotropic stress is turned on by the presence of
$\tau_Y$, $\tau_Z$. $w_Y$ and $w_Z$.  In the scalar sector, the
velocity $\delta\mathfrak{U}^\m $  and the anisotropic stress perturbations  $\Pi^\mu_\nu$ 
can be written in terms of two extra scalars $v$ and $\Xi$~\footnote{Note that the dimension of these two extra scalars is $[v]=-1$ and $[\Xi]=2$.}, in
addition to $\delta \rho$
\be
\begin{split}
&\delta \mathfrak{U}_\mu=(a\;\Psi,\;   a \, \partial_i  v),\qquad
\Pi^\mu_\nu\equiv\;
(3\;\partial^2  \, \delta^\mu_i\; \delta^i_\nu - \delta^\mu_i
\, \partial^i \, \delta_\nu^j \partial_j)\; \Xi \, , \\
& \Xi \equiv -2 \, a^2 \, M_2 \, \pi_{L\, gi}=2 \left[\Phi-\Psi+2{\cal H}\, (F-B')+(F-B')'\right]\, ,
\end{split}
\ee
with
\be
\begin{split}
&\delta \rho=-\bar \rho (1+w)(3 \Phi + k^2 \pi_{L,\, gi}) + a^{-4}\,\bar\varphi' \,\delta\,\sigma \,,\\
&v=-\pi_{L,\, gi}' +\frac{M_1 \, a^2 }{6 \,{\cal H}^2\, (1+w)}\left[ \frac{\pi_0}{\bar\varphi'}-\pi_{L , \, gi}{}' + (F-B')\right]+(F-B')\, . 
\end{split}
\ee
Using this parameterization, ${\cal R}$,  $\zeta$ and $\sigma$ can
be easily written in terms of $\pi_L$, $\pi_0$ and their derivative
w.r.t. conformal time.  Namely, with the suffix $gi$ understood, we have
\be
\label{R}
\begin{split}
{\cal R}=& \epsilon\, (1+c_1^2)\, \left(1-2 \, c_0^2 \,c_b^2\right)\,{\cal H}^2 \, \pi_L+c_1^2 \, {\cal H}\, \left[-1+3 \, \epsilon\, (1+c_1^2) \, \frac{{\cal H}^2}{k^2}\right]\, \frac{\pi_0}{\bar\varphi'}\\
&+(1+c_1^2)\, \left[1-3 \, \epsilon\, (1+c_1^2) \, \frac{{\cal H}^2}{k^2}\right]\,{\cal H}\, \pi_L'-2 \, \epsilon\, (1+c_1^2)\, \frac{{\cal H}^2}{k^2}\, \frac{\pi_0'}{\bar\varphi'}\,,
\end{split}
\ee
\be 
\label{zeta}
\begin{split}
\zeta=&\frac{k^2}{3} \, (1-2\, c_0^2\, c_b^2)\, \pi_L- \frac{2}{3}\, \epsilon\, c_0^2 \, c_1^2 \, {\cal H}\, \frac{\pi_0}{\bar\varphi'}\\
&+\frac{2}{3}\, \epsilon \, c_0^2 \, {\cal H}\, (1+c_1^2)\, \pi_L'-\frac{2}{3}\, c_0^2 \, \frac{\pi_0'}{\bar\varphi'}\,,
\end{split}
\ee
\be 
\label{sigma}
\delta\sigma=-2 \, a^4 \, \bar{\rho} \, \bar\varphi'^{-1}\,\left(\zeta-\zeta_n\right)\, \epsilon\,.
\ee
The above expressions are valid at the linear order in the slow-roll parameters.

\section{Canonical Transformation}
\label{can-app}

In the UV (large $k$) the Lagrangian (\ref{lagrangian}) becomes
\be
L_2^{\text{(UV})} = \ha \Pi'{}^t \Pi' - \Pi^t \, {\cal D}^{\text{UV}} \,
\Pi' - \ha \Pi^t\, {\cal M}^{\text{UV}} \, \Pi  \;
\label{lagrangianUV}
\ee
with 
\be
{\cal D} \to {\cal D}^{\text{UV}} \equiv {\cal D} \, , \qquad  {\cal M} \to {\cal M}^{\text{UV}} = \begin{pmatrix}
k^2\, \lambda_L{}^2  & 0\\
0 & k^2\, \lambda_0{}^2
\end{pmatrix}
\ee
The first step is to find the Hamiltonian density corresponding to
(\ref{lagrangian}), which reads
\be
\label{hamiltonian}
{\cal H}= \frac{1}{2}\, P^t\, P-P^t\, {\cal D}\, \Pi
+\frac{1}{2}\,\Pi^t\,\left({\cal M}-{{\cal D}^{\text{UV}}}^2\right)\,\Pi\,,
\ee
where $P$ is the conjugate momentum of $\Pi$
\be 
P= \Pi'+{{\cal D}^{\text{UV}}}\, \Pi\,.
\ee
The decoupled system can be obtained by applying a canonical transformation of the form:
\be
\begin{split}
\label{can_tr}
&\Pi= A \, \tilde \Pi+ B\, \tilde P\,, \\
& P= J \, \tilde P+ C\, \tilde \Pi\,. \\
\end{split} 
\ee 
Using (\ref{par}), imposing that the transformation is canonical with the new ${\cal D}_c$ vanishing, the four matrices $A$, $B$, $J$ and $C$ can be taken of the following form 
\be
 A=
 \begin{pmatrix}
1 & 0 \\
0 & 1

\end{pmatrix} \,, \;\;\; 
J=\frac{1}{2}\left(1+\frac{\Delta \lambda}{\Delta}\right)
 \begin{pmatrix}
1 & 0 \\
0 & 1
\end{pmatrix}\,, \;\;\;
B=-2\,\frac{d}{k\,\Delta}
\begin{pmatrix}
0 & 1 \\
1 & 0
\end{pmatrix}\,, \;\;\;
C=-k\,\frac{\Delta \lambda-\Delta}{4\,d}
\begin{pmatrix}
0 & 1 \\
1 & 0
\end{pmatrix}\,,
\ee
where $\Delta \lambda$ and $\Delta$ are given by 
\be
\Delta \lambda= \lambda_0^2-\lambda_L^2\,, \qquad \Delta =\frac{\left( 4\, c_0^4\, c_L^4+ c_1^4 \, \left[1+8 \, c_0^2 (c_L^2-c_b^2)\right]+4\, c_0^2\, c_1^2\, \left[c_L^2+4\, c_0^2\, c_b^2\, (c_L^2-c_b^2)\right]\right)^{\frac{1}{2}}}{2\, c_0^2 \, (1+c_1^2)} \,.
\ee
The transformed Hamiltonian reads
\be
{\cal H}_{\text{new}}= \frac{1}{2}\, \tilde{P}^t \, {\cal K}_{\text{new}} \, \tilde{P}
+\frac{1}{2}\,\tilde{\Pi}^t\,{\cal M}_{\text{new}} \,\tilde{\Pi} \, ,
\label{newH}
\ee
with
\be
{\cal K}_{\text{new}}=A+B\,(C+{\cal D}) \equiv \begin{pmatrix} {\cal
    K}_{1 \text{new}} & 0 \\ 0 &  {\cal
    K}_{2 \text{new} }\end{pmatrix}  \,,\qquad {\cal
  M}_{\text{new}}=-B^{-1}\, (C-{\cal D})  \equiv \begin{pmatrix} {\cal
    M}_1 & 0 \\ 0 &  {\cal
    M}_2 \end{pmatrix}  \,;
\ee
both ${\cal K}_{\text{new}}$ and  ${\cal
  M}_{\text{new}}$ are diagonal and thus, for fixed $k$, (\ref{newH})
describes two uncoupled harmonic oscillators with frequencies $k^2 \,
c_{s1}^2$ and  $k^2 \,
c_{s2}^2$, where
\be\label{css}
 c_{s 1}^2=\frac{{\cal M}_1 \,  {\cal K}_{1 \text{new}}}{ k^2 } =
   d^2+\lambda_L^2+\frac{\Delta \lambda}{2}+\frac{\Delta}{2}
   \,,\qquad  c_{s2}^2=\frac{{\cal M}_2 \,  {\cal K}_{2
       \text{new}}}{ k^2 }  =2 \, d^2+\lambda_L^2+\frac{\Delta
   \lambda}{2}
   -\frac{\Delta}{2} \,.
\ee
Quantization of (\ref{newH}) is straightforward: the Bunch-Davies
vacuum is the state $|0 \rangle$ of the Fock space corresponding to the following
field operators
\be
\label{diagonal_fields}
 \tilde \Pi
 =\begin{pmatrix}
 \tilde \Pi_1 \\
 \tilde \Pi_2
 \end{pmatrix}
 =
 \begin{pmatrix}
 A_k^{(1)}\, a_k{}^{(1)}\, e^{-i\, k\, c_{s1}\, t}+A_k^{(1)\,*}\, a_k{}^{(1)\,\dagger}\, e^{i\, k\, c_{s1}\, t} \\
 A_k^{(2)}\, a_k{}^{(2)}\, e^{-i\, k\, c_{s2}\, t}+A_k^{(2)\,*}\, a_k{}^{(2)\,\dagger}\, e^{i\, k\, c_{s2}\, t} 
\end{pmatrix}\,, \qquad A_k^{(n)}=k^{-\frac{1}{2}}\, \left(\frac{{\cal
      K}_n}{c_{s n}}\right)^{\frac{1}{2}} 
\ee
with  $a_k{}^{(1/2)\,\dagger}$ and $a_k{}^{(1/2)}$ standard creation and
  annihilation operators. The fields satisfies the following canonical
  commutation relations
\be
\left[\tilde \Pi_{m}(t, \textbf{x})\,,\;\tilde \Pi_{n}'(t,
  \textbf{y})\right]=i\, {\cal K}_m\, \delta^{(3)} \left(
  \textbf{x}-\textbf{y}\right)\,\delta_{mn}\,, \qquad m,n =1,2 \, .
\ee
By using (\ref{can_tr}), one can express $\Pi_L$ and $\Pi_0$ in terms
of the creation and annihilation operators 
\be
\Pi_L=\sum_{j=1}^2 C_L^{(j)}\,a_k^{(j)} \,e^{-i\, k\, c_{s\,j}\,t}
+C_L^{(j)\,*}\,a_k^{(j)\, \dagger} \,e^{i\, k\, c_{s\,j}\,t}
, ,
\label{piLexp}
\ee
\be
\Pi_0=\sum_{j=1}^2 C_0^{(j)}\,a_k^{(j)} \,e^{-i\, k\, c_{s\,j}\,t}
+C_0^{(j)\,*}\,a_k^{(j)\, \dagger} \,e^{i\, k\, c_{s\,j}\,t}\,,
\label{pi0exp}
\ee
where  $C_{L/0}^{(j)}$ read
\be
C_L^{(1)}=k^{-\frac{1}{2}}\, \left(\frac{{\cal K}_{1 \text{new}}}{
    c_{s1}}\right)^{\frac{1}{2}}\,,\qquad C_L^{(2)}=i \,
k^{-\frac{1}{2}}\,\frac{2\,d}{\Delta}\, \left(\frac{ c_{s2}}{{\cal
      K}_{2 \text{new}}}\right)^{\frac{1}{2}} \,, 
\ee
\be
C_0^{(1)}=i \,k^{-\frac{1}{2}}\,\frac{2\,d}{\Delta}\, \left(\frac{c_{s1}}{{\cal K}_{1 \text{new}}}\right)^{\frac{1}{2}}\,,\qquad
C_0^{(2)}=k^{-\frac{1}{2}}\, \left(\frac{{\cal K}_{2 \text{new}}}{c_{s2}}\right)^{\frac{1}{2}}\,.
\ee
By using (\ref{piLexp}-\ref{pi0exp}), one can compute the free-field
(Gaussian) average of any operator expressed in terms of
$\pi_L$ and $\pi_0$.
In order to simplify as much as possible the expression of power spectra, it is rather useful to rewrite all the parameters of interest in terms of the two ``diagonal'' sound speeds $c_{s1/2}$, $c_b$ and $c_L$ defined in (\ref{cl}). We get
\be 
\begin{split}
c_0^2 &=\frac{\left(c_L^2-c_{s1}^2\right) \left(c_L^2-c_{s2}^2\right)}{2 c_b^4 \left(c_L^2-c_{s1}^2\right)-2 c_{s2}^2 \left(-2 c_b^2 c_{s1}^2+c_b^4+c_L^2
   c_{s1}^2\right)}\,,\\
c_1^2 &=\frac{\left(c_L^2-c_{s1}^2\right) \left(c_L^2-c_{s2}^2\right)}{-2 c_b^2 c_L^2+c_b^4+c_L^2 \left(c_{s1}^2+c_{s2}^2\right)-c_{s1}^2 c_{s2}^2}\,,\\
d^2&= \frac{\left(c_b^2-c_{s1}^2\right){}^2 \left(c_b^2-c_{s2}^2\right){}^2 \,\left(c_L^2-c_{s1}^2\right) \left(c_L^2-c_{s2}^2\right)}{4 \left[c_b^4
   \left(c_L^2-c_{s1}^2\right) -c_{s2}^2 \left(-2 c_b^2 c_{s1}^2+c_b^4+c_L^2 c_{s1}^2\right)\right] \left(-2 c_b^2 c_L^2+c_b^4+c_L^2
   \left(c_{s1}^2+c_{s2}^2\right)-c_{s1}^2 c_{s2}^2\right)}\,,\\
{\cal K}_1 &=\frac{\left(c_b^2-c_{s1}^2\right){}^2 \left(c_L^2-c_{s2}^2\right)}{\left(c_{s1}^2-c_{s2}^2\right) \left(-2 \,c_b^2 c_L^2+c_b^4+c_L^2
   \left(c_{s1}^2+c_{s2}^2\right)-c_{s1}^2 c_{s2}^2\right)}\,,\\
{\cal K}_2 &= \frac{c_{s2}^2 \left(c_b^2-c_{s1}^2\right){}^2 \left(c_{s2}^2-c_L^2\right)}{\left(c_{s1}^2-c_{s2}^2\right) \left(c_b^4
   \left(c_L^2-c_{s1}^2\right)-c_{s2}^2 \left(-2 c_b^2 c_{s1}^2+c_b^4+c_L^2 c_{s1}^2\right)\right)}\,,\\
\Delta &= c_{s2}^2-c_{s1}^2\,.
\end{split}
\ee
Finally 
\be
\mid C_{L}^{(1)}\mid ^2 \, k=\frac{\left(c_b^2-c_{s1}^2\right){}^2 \mid c_L^2-c_{s2}^2\mid}{c_{s1} \mid c_{s1}^2-c_{s2}^2\mid \mid-2 c_b^2 c_L^2+c_b^4+c_L^2
   \left(c_{s1}^2+c_{s2}^2\right)-c_{s1}^2 c_{s2}^2\mid}
\ee
\be
 \mid C_{L}^{(2)}\mid ^2\, k =\frac{\left(c_b^2-c_{s2}^2\right){}^2 \mid c_L^2-c_{s1}^2\mid}{ c_{s2} \mid c_{s1}^2-c_{s2}^2\mid \mid-2 c_b^2 c_L^2+c_b^4+c_L^2
   \left(c_{s1}^2+c_{s2}^2\right)-c_{s1}^2 c_{s2}^2\mid}
\ee
\be
 \mid C_{0}^{(1)}\mid ^2 k=\frac{c_{s1} \left(c_b^2-c_{s2}^2\right){}^2 \mid c_{s1}^2-c_L^2\mid }{ \mid c_{s1}^2-c_{s2}^2\mid  \, \mid c_b^4
   \left(c_L^2-c_{s1}^2\right)-c_{s2}^2 \left(-2 c_b^2 c_{s1}^2+c_b^4+c_L^2 c_{s1}^2\right)\mid}
\ee
\be 
\mid C_{0}^{(2)}\mid ^2 k=\frac{c_{s2} \left(c_b^2-c_{s1}^2\right){}^2 \mid c_{s2}^2-c_L^2\mid }{ \mid c_{s1}^2-c_{s2}^2\mid \mid c_b^4
   \left(c_L^2-c_{s1}^2\right)-c_{s2}^2 \left(-2 c_b^2 c_{s1}^2+c_b^4+c_L^2 c_{s1}^2\right)\mid }
\ee
Note that under the exchange of $c_{s 1}
\leftrightarrow c_{s2} $ we have that $ C_{L/0}^{(1)} \leftrightarrow
C_{L/0}^{(2)}$. As a consequence, all the power spectra will have the same property. Such a symmetry simply reflects the conventional choice $c_{s2}<c_{s1}$ or $c_{s1}<c_{s2}$.
\\
Finally, note that in the two analytic cases $c_b=0,\,-1$, ${\cal C}_{L/0}$ sub-horizon coefficients are easily transmitted on superhorizon scales, and eq. (\ref{PZ}) can be obtained 
considering that $\zeta_n$ and ${\cal R}_{\pi_0}$ classical solutions reduce to
\be
\label{super-rel}
\begin{split}
&\zeta_n{}^{(j)} \to - \frac{{\cal C}_{L}^{(j)}\, H}{2\,\sqrt{1+c_1^2}\, c_{sj}^2\, M_p\, \epsilon^{\frac{1}{2}}}\, k^{-1}\,,\\
&{\cal R}_{\pi_0}{}^{(j)} \to \frac{i\,{\cal C}_{0}^{(j)}\, H}{2\,\sqrt{2}\, c_0\,c_{sj}\, M_p \, \epsilon^{\frac{1}{2}}} \, k^{-1}\,. 
\end{split} 
\ee

\end{appendix}

\bibliographystyle{unsrt}  
\bibliography{infl}

\end{document}